 \documentclass{aa}
\usepackage{graphicx}
 \usepackage{epsfig,times}
 \usepackage{color}

\begin{document}

\title{Formation of protostellar jets - effects of 
       magnetic diffusion}
%
%
\author{
        Christian Fendt
          \inst{1,2},
        Miljenko \v{C}emelji\'{c}
          \inst{2}
       }
\offprints{Ch.~Fendt}
\institute{
              Universit\"at Potsdam, Institut f\"ur Physik,
              Am Neuen Palais 10, 14469 Potsdam, Germany
          \and
              Astrophysikalisches Institut Potsdam,
              An der Sternwarte 16, 14482 Potsdam,
              Germany;
              \email{cfendt@aip.de, cemeljic@aip.de}
          }
\date{Received 24 May 2002 / 24 September 2002 }
\titlerunning{Diffusive protostellar jet formation}
\authorrunning{Fendt \& \v{C}emelji\'{c} }
\abstract{
Protostellar jets most probably originate in turbulent accretion disks
surrounding young stellar objects.
We investigate the evolution of a disk wind into a collimated jet
under the influence of magnetic diffusivity, assuming that the turbulent
pattern in the disk will also enter the disk corona and the jet.
Using the ZEUS-3D code in the axisymmetry option we solve the
time-dependent resistive MHD equations for a model setup of a central
star surrounded by an accretion disk.
The disk is taken as a time-independent boundary condition for the
mass flow rate and the magnetic flux distribution.
We derive analytical estimates for the magnitude of magnetic diffusion 
in a protostellar jet connecting our results to earlier work in the 
limit of ideal MHD.
We find that the diffusive jets propagate slower into the ambient
medium, most probably due to the lower mass flow rate in the axial
direction.
Close to the star we find that a quasi stationary state evolves after
several hundred (weak diffusion) or thousand (strong diffusion) 
disk rotations.
Magnetic diffusivity affects the protostellar jet structure as follows.
The jet poloidal magnetic field becomes de-collimated.
The jet velocity increases with increasing diffusivity, while
the degree of collimation for the hydrodynamic flow remains more or
less the same.
We suggest that the mass flux is a proper tracer for the degree of
jet collimation and find indications of a critical value for the
magnetic diffusivity above which the jet collimation is only weak.
We finally develop a self-consistent picture in which all these
effects can be explained in the framework of the Lorentz force.
}

\maketitle
\def\ri{r_{\rm i}}
\def\ro{r_{\rm out}}
\def\rs{r_{\star}}
\def\mj{\dot{M}_{\rm jet}}
\def\ms{\dot{M}_{\star}}
\def\vk{v_{\rm K}}
\def\rj{R_{\rm jet}}
\def\rk{R_{\kappa}}
\def\bp{B_{\rm p}}
\def\bh{B_{\phi}}
\def\jp{\vec{B}_{\rm P}}
\def\cm{{\rm cm}}
\def\msun{{\rm M}_{\sun}}
\def\rsun{{\rm R}_{\sun}}

\section{Introduction}
Observations of young stellar objects (YSOs) have revealed two main 
features during the phase of star formation --
the presence of accretion disks and energetic outflows,
often observed as bipolar jets 
(Mundt et al. \cite{mundt}, 
 Lada \cite{lada}, 
 Ray et al. \cite{ray}).
Images and spectra show that these flows are of high-velocity 
($\simeq 300 {\rm km s^{-1}}$)
and well collimated (opening angle $<10\degr$ on scales of 1000 AU). 
The data also suggest that the jet collimation must be achieved already
close to the central source, at distances $\leq$100 AU.
The jet mass outflow rates are $\sim$ $10^{-10}-10^{-8}\msun/{\rm yr}$
and typically a factor 10-100 smaller than the disk accretion rates.

Besides the special case of protostellar jets, 
all astrophysical jets detected so far seem to be attached to objects where an
accretion disk is indicated to be present.
In particular, this holds for jets observed in radio loud active galactic
nuclei and quasars 
(Bridle \& Perley \cite{bridle}), 
highly energetic galactic objects as Sco X--1 
(Padman et al.\cite{padman}), 
and microquasars (Mirabel et al.\cite{mirabel}).

Therefore, 
the similarities between jets from the different sources imply that the
basic mechanism for jet formation should be the same.
For protostellar jets, the observed mass and momentum fluxes exclude
the possibility of a thermally or radiation pressure driven wind. 
The observed fluxes are much higher than the protostar could provide
(DeCampli \cite{decampli}, K\"{o}nigl \cite{koenigl86}). 
The conclusion is that it is the magnetic field which is responsible
for protostellar jet formation -- acceleration and collimation of the
initial stellar or disk wind
(Pudritz \& Norman \cite{pudritz}, 
 Camenzind \cite{camenzind}). 
This magnetic field can be generated by some dynamo process either in 
the central young star itself, or
the surrounding accretion disk, or it could be provided by the interstellar 
medium as a ``fossil'' field.
What kind of mechanism turns the in-flowing matter of the accretion disk into
an outflow from the disk (or the star) is still not really known, although
it seems to be clear that magnetic fields play a major role 
(Ferreira \cite{ferreira}).

Examples of time--dependent MHD simulations of the jet formation 
include models of collimating disk winds, where this disk is taken
as a boundary condition for the outflow 
(Ouyed \& Pudritz \cite{ouyed}, hereafter OP97),
models that consider the interaction of a stellar dipolar magnetosphere
with the accretion disk as well as the disk structure itself
(Miller \& Stone \cite{miller97}),
or some combination of both approaches
(Fendt \& Elstner \cite{fendt99}, \cite{fendt00}, hereafter FE00).

In this paper, we are interested in a time--dependent simulation of the
{\it resistive} MHD equations for the corona above the accretion disk around a 
young star. 
The underlying disk provides a fixed boundary. 
The presence of a large scale poloidal magnetic field provided by the disk
is assumed. 
After a brief theoretical introduction in \S 2, we describe our model and the 
numerical approach in \S 3, with emphasis on the effect of magnetic diffusion.
For the numerical approach, we introduced magnetic diffusion in the
ZEUS-3D MHD code. Tests of our code are included in Appendix. 
The results for protostellar jets are discussed in \S 4. 
We compare our diffusive MHD simulations of jet formation with the 
non--diffusive case. 

\section{Magnetic jets from accretion disks}
Despite a tremendous amount of work concerning the formation of magnetic jets 
(e.g. Blandford \& Payne \cite{blandford}, Pudritz \& Norman \cite{pudritz},
Lovelace et al.\cite{lovelace}, 
Heyvaerts \& Norman \cite{heyvaerts}, 
Pelletier \& Pudritz \cite{pelletier}, 
Li et al. \cite{li92}, 
Contopoulos\cite{contopoulos}, 
Fendt et al. \cite{fendt95}, 
Kudoh \& Shibata \cite{kudoh97a}, \cite{kudoh97b}, 
Fendt \& Memola \cite{fendt01})
the mechanism which actually launches the jet from the disk remains unclear.
Most of the papers dealing with the theory of magnetized accretion disks 
driving jets have been following the principal approach of Blandford \& Payne
assuming stationarity, axisymmetry and self-similarity
(K\"onigl \cite{koenigl89}, Wardle \& K\"onigl \cite{wardle}, 
Li \cite{li95}, Ferreira \cite{ferreira}). 
After all, it is clear now that the launching of a jet from the accreting 
disk can be described as a purely {\em magnetic} process.
Ferreira (\cite{ferreira}) 
has derived trans--Alfv\'{e}nic, stationary self-similar jet
solutions with a smooth transition from a resistive disk, where the
Lorentz forces are actually responsible for lifting the accreting gas
in vertical direction.

Progress has been achieved, too, 
in the simulation of the time-dependent MHD jet formation from accretion disks. 
Two major, distinct approaches in order to deal with the complexity of the
jet formation process have been undertaken so far.
One approach is to take the rotating disk as a fixed boundary condition
for the simulation of the jet.
Depending on the choice of the initial setup (magnetic field, density and pressure) 
and the choice of the gravitational potential, 
the numerical results differ in the degree of collimation and the velocity of 
the resulting jet flow. 
From an initially split-monopole magnetic field configuration,
collimated, non-stationary outflows were obtained 
(Ustyugova et al.\cite{ustyugova}).
In this simulation non-equilibrium initial conditions and a softened 
gravitational potential were used. 
For the same configuration, but for a stronger magnetic field, 
Romanova et al. (\cite{romanova}) again 
obtained stationary but only weakly collimated flows. 
Ouyed \& Pudritz (\cite{ouyed}) studied the jet formation embedded in a disk corona 
initially
in hydrostatic equilibrium and in pressure balance with the disk surface.
Essentially, after 400 disk rotations, a stationary collimated jet flow 
emerges.
Similar results were obtained by another recent study taking into account 
also the time-dependent behavior of the disk boundary condition (namely the
field inclination) due to the evolution of the disk wind 
(Krasnopolsky et al.\cite{krasno}).
The main advantage of the approach of a fixed disk boundary condition
is the numerical stability of simulation over a long time scale.
The jet launching itself -- the process of diverting accreting matter in the
disk into an outflow -- cannot be treated this way.

The other approach was therefore to include the simulation of the disk
structure in the simulation.
The first step in this direction was made by Uchida \& Shibata 
(\cite{uchida}) and Shibata \& Uchida (\cite{shibata85}, \cite{shibata86}) 
in their pioneering work considering time-dependent jet formation.
Essentially, the authors show that the magnetic twist of the magnetic field
induced by the rotation of the disk gives rise to Lorentz forces pushing
the disk material upwards. 
The back-reaction of the magnetic field on the disk (magnetic braking) may
lead to a sub-Keplerian disk rotation.
These results have been confirmed by Stone \& Norman (\cite{stone94}).
MHD simulations considering the diffusive accretion disk in interaction with a 
stellar dipolar magnetosphere 
reveal the collapse of the inner disk after a few rotations
(Hayashi et al. \cite{hayashi}, Goodson et al. \cite{goodson}, 
Miller \& Stone \cite{miller97}). 
At the same time, episodic ejection of plasmoids are generated in outer parts
of the disk wind.
A collimated axial jet feature is observed.
However, probably because of numerical reasons, 
these simulations could be performed only for a few or tens of Keplerian periods 
of
the inner disk.
Therefore, the results may depend strongly on the initial setup.
The assumption of constant diffusivity in the disk and the corona is probably
not very realistic.

The long-term evolution of such models has been investigated by FE00,
however without treating the disk structure in the simulation.
They find that the axial jet feature observed by Goodson et al.~(\cite{goodson}) 
disappears on
longer time-scales. A two-component quasi-stationary outflow (from disk and 
star)
evolves after thousands of rotational periods.
This flow is un-collimated on the spatial scales investigated, in agreement with
the observations indicating a jet radius 100 times larger then the grid size of 
the
numerical simulations discussed above.

Recent MHD simulations of the jet formation from accretion disks
by  Matsumoto et al. (\cite{matsumoto}), 
Kudoh et al. (\cite{kudoh98}, \cite{kudoh02a}, \cite{kudoh02b} ),
and
Kato et al. (\cite{kato})
investigate the disk-jet {\em interrelation} and may probably explain
the time-dependent ejection mechanism in the jets.
However, as we have pointed out above for the model setup of a magnetic
dipole surrounded by a disk,
these simulations also were carried out for several rotation periods 
only.
The question remains as to how the system under consideration behaves
on a long time scale.
Further, the simulations apply the approach of {\em ideal} MHD,
an approximation, which is most probably not strictly valid for
accretion disks, 
especially for protostellar accretion disks.
In this respect, 
the work by Kuwabara et al. (\cite{kuwabara}) is of particular
interest as the authors extend the ideal MHD approach and include 
{\em resistivity} in the jet formation process.
Comparing their simulations to the close environment of a supermassive 
black hole, Kuwabara et al. derive a {\em critical value} for the strength 
of magnetic diffusion. 
A normalized magnetic diffusivity below $0.05$ may explain the observed
activity in active galactic nuclei. 
In this case, the mass accretion and jet launching takes place
intermittently.
The paper also demonstrates the difficulty of carrying out such simulations,
as gravity has been treated applying a softened potential. 
On the other hand, this simplification allows for a simulation lasting
up to 5 rotations of the accretion torus.

Again, we note that it is just the limitation in the time evolution
which lead us to the decision to take into account the accretion disk only 
as a boundary condition. 
In this paper, we are mainly interested in the jet formation process
(acceleration and collimation) and 
not in the jet launching mechanism from the accretion disk.
In particular, we think it is essential to investigate whether a jet
actually survives also on long time scales.

\section{Model setup and equations }
The paradigm accepted in this paper is the jet launched from a diffusive, 
turbulent accretion disk around a young stellar object. 
We may expect that the turbulence pattern in the disk may also enter the 
disk corona and the jet,
and that the jet flow itself is subject to turbulent diffusion.
We will discuss this idea of a diffusive protostellar jet
below.
Despite the fact that we take into account the effect of magnetic diffusion
for the jet formation, our model setup is similar to the models in OP97 and FE00.

\subsection{Resistive MHD equations}
In order to model the time-dependent evolution of jet formation, the set
of resistive MHD equations to be solved is

\begin{equation}
{\frac{\partial \rho}{\partial t}} + \nabla \cdot (\rho \vec{v} ) = 0
\end{equation}

\begin{equation}
\rho \left[ {\frac{\partial\vec{u} }{\partial t}}
+ \left(\vec{v} \cdot \nabla\right)\vec{v} \right]
+ \nabla (p+p_A) +
 \rho\nabla\Phi - \frac{\vec{j} \times \vec{B}}{c} = 0
\end{equation}

\begin{equation}
{\frac{\partial\vec{B} }{\partial t}}
- \nabla \times \left(\vec{v} \times \vec{B}
-{\frac{4\pi}{c}} \eta\vec{j}\right)= 0
\end{equation}

\begin{equation}
\rho \left[ {\frac{\partial e}{\partial t}}
+ \left(\vec{v} \cdot \nabla\right)e \right]
+ (p+p_A) (\nabla \cdot\vec{v} )
- {\frac{4\pi}{c^2}}\eta \vec{j}^2= 0
\end{equation}

\begin{equation}
\nabla \cdot\vec{B} = 0
\end{equation}

\begin{equation}
\frac{4\pi}{c} \vec{j} = \nabla \times \vec{B}
\ .
\end{equation}
For the gas law we apply a polytropic equation of state,
$p=K\rho^\gamma$ with a polytropic index $\gamma=5/3$.
Hence, we do not solve the energy  equation (4).
Instead, the internal energy of the system is defined with
$ e=p/(\gamma-1)$. 
Such a simplification is not expected to affect the result of our simulations
much, 
as the resistive dissipation term in the energy equation is negligible compared
to the other terms because of the factor $1/c^2$ (see also 
Miller \& Stone \cite{miller97}).
The magnetic diffusivity is denoted by the variable $\eta$ (see Sect.\,3.3).
Additional to the hydrostatic pressure $p$, an Alfv\'{e}nic turbulent pressure 
$p_{\rm A}\equiv p/\beta_{\rm T}$ with $\beta_{\rm T} = {\rm const.}$
(see OP97, FE00) is included.
Alfv\'{e}n waves from the highly turbulent accretion disk are expected to
propagate into the disk corona, 
providing the perturbations for some degree of turbulent motion also in the jet.
As shown by OP97 the additional Alfv\'{e}nic turbulent pressure is able to support a
{\em cold} corona above a protostellar accretion disk as suggested by the
observations.

We solve the equations above using the ZEUS-3D code (Stone \& Norman 
\cite{stone92a},\cite{stone92b}) 
in the 2D-axisymmetry option for cylindrical coordinates $(r,\phi,z)$. 
We apply a point mass gravitational potential 
$\Phi = - 1/\sqrt{r^2 + z^2}$ located in the origin.
A finite physical magnetic resistivity is added to the original
ZEUS-3D ideal MHD code.
Tests of our now diffusive ZEUS code are presented in the Appendix.

For our computations we normalize the variables to their value measured
at the inner disk radius $r_{\rm i}$ (see OP97, FE00),
e.g. $\rho \rightarrow \rho /\rho_{\rm i}$.
The subscript `i' assigns that the variables are taken at this radius.
The time is measured in units of a Keplerian rotation at the inner disk 
radius.
The normalized equation of motion eventually being solved with the
code is
\begin{equation}
\frac{\partial\vec{v}' }{\partial t'}
+ \left(\vec{v}' \cdot \nabla'\right)\vec{v}' =
\frac{2 \,\vec{j}' \times \vec{B}' }{\delta_i\,\beta_i\,\rho'}
- \frac{\nabla' (p' + p_{\rm A}')}{\delta_i \,\rho'} - \nabla'\Phi'\,.
\end{equation}
The coefficients
$\beta_i \equiv 8 \pi p_i / B_i^2 $ and
$\delta_i \equiv \rho_i v^2_{K,i} / p_i $ with
the Keplerian speed $v_{K,i} \equiv \sqrt{GM/r_i} $,
correspond to the plasma beta and the Mach number of the
gas at the inner disk radius.
For a ``cold'' corona with $p_{\rm A}' > 0$, it follows
$\beta_T = 1 / (\delta_i (\gamma - 1) / \gamma - 1)$.
In the following we will omit the primes and will discuss only
normalized variables if not explicitly declared otherwise.

\begin{figure}
\centering
\includegraphics[bb= 18 144 592 518,width=9cm]{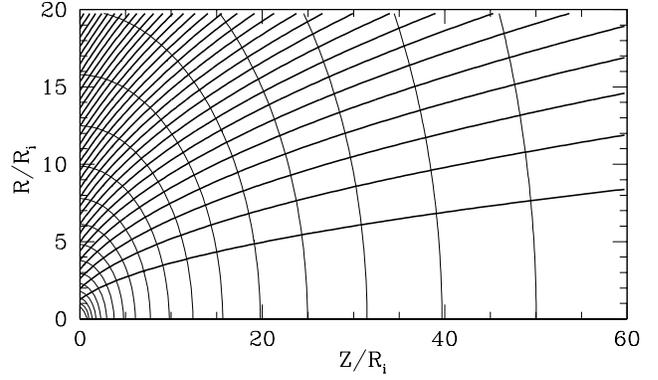}
 \caption{
Initial setup for the jet simulation.
Shown is the part of the computational box close to the origin
(the part of the ``inner jet'').
The initial hydrostatic density distribution is indicated by the
{\it thin} concentric isocontours.
{\it Thick} lines denote the initial poloidal field lines of a force-free
potential field.
}
\end{figure}

\subsection{Initial and boundary conditions}
As the numerical simulation of a magnetized disk is still a difficult task
and is not yet fully resolved,
we chose to study the formation of the jet flow independent of the
evolution of the accretion disk. 
A precise {\em disk}-wind theory would predict the amount of angular momentum and
energy carried away from the disk. 
Here, we prescribe the disk as a fixed, time-independent boundary condition
for the jet.

One may question the combination of a magnetically diffusive disk corona
and a steady-state disk magnetic field distribution.
Naturally, the time scales of the disk evolution are definitely longer than 
those for the jet flow 
(except for the unknown mechanism responsible for the jet knots).
Also, for a disk jet, the disk is acting as a source for magnetic flux.
This can be achieved either by a dynamo process working in the disk generating 
the magnetic field or just by advection of the interstellar magnetic field by
the disk towards the central star.
The time scale for both processes is longer than the jet time scale and, thus,
we may safely assume a fixed magnetic flux as accretion disk boundary condition.

\begin{figure*}
\centering
\includegraphics[width=19cm]{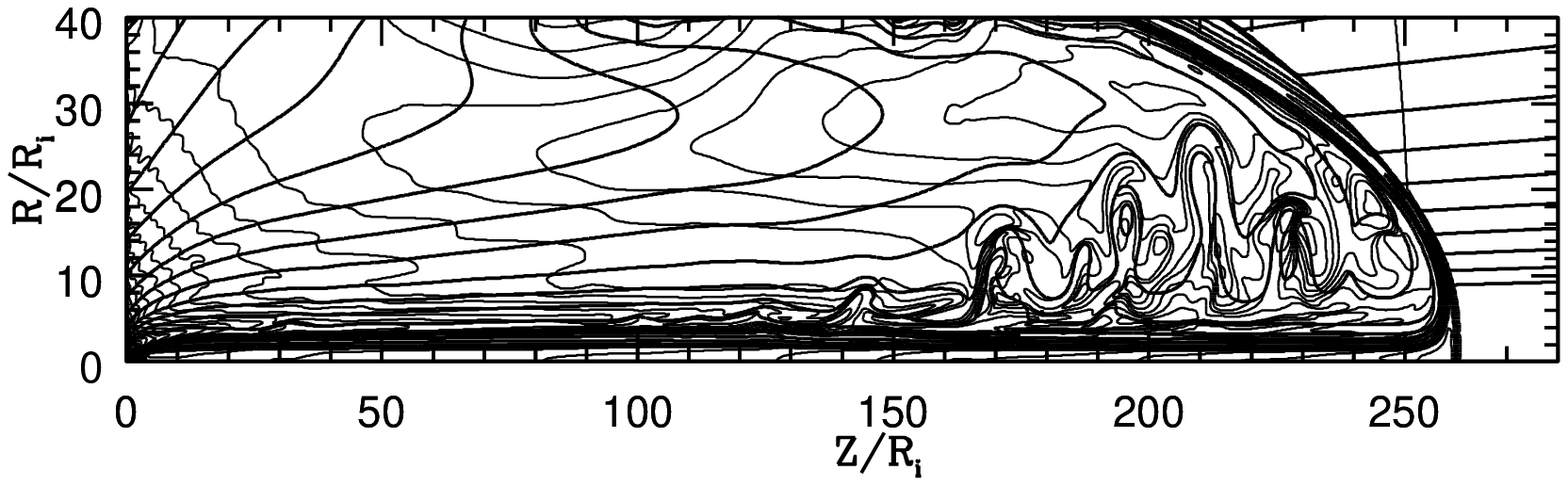}
\includegraphics[bb=18 144 592 298,width=19cm]{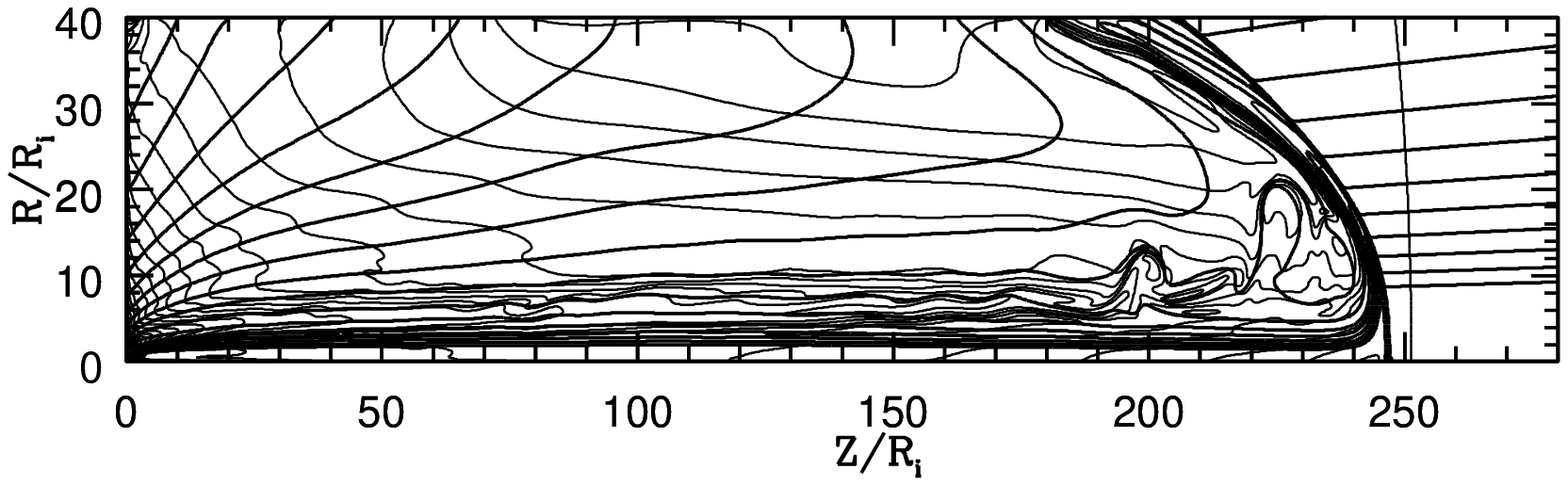}
\includegraphics[bb=18 244 592 298,width=19cm]{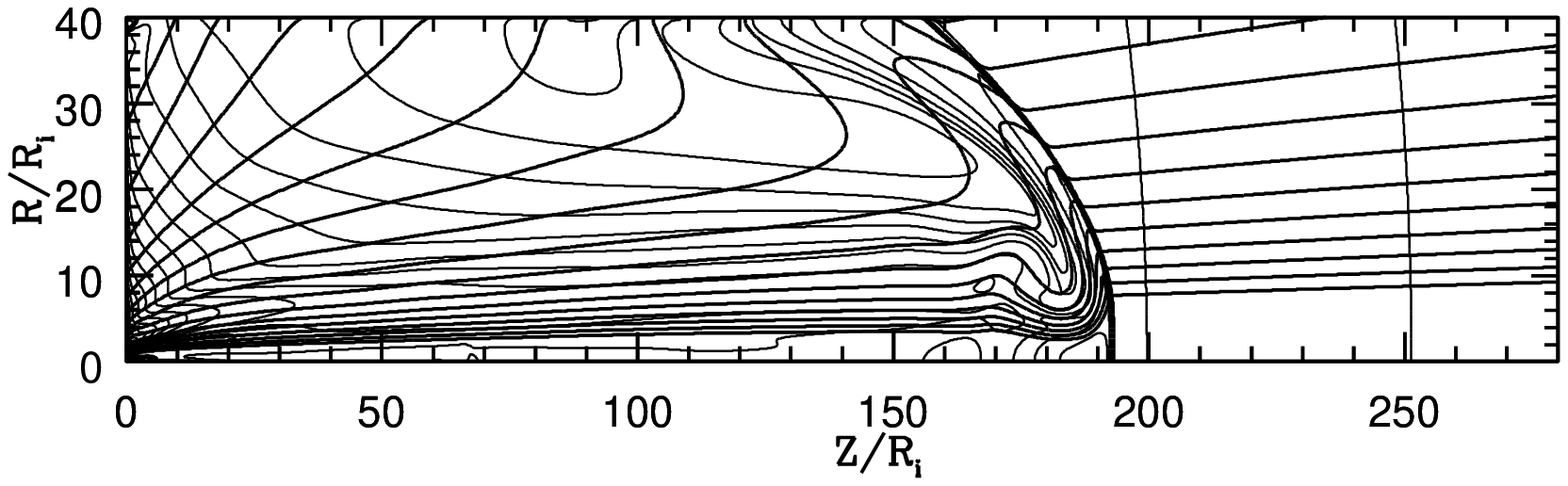}
\vspace{-8cm}
\caption{Global evolution of the jet on a grid of 
($z\times r$)=(280$\times$ 40)$r_{\rm i}$ with a resolution 
of 900$\times$200 elements. 
Shown is the state of evolution after $t=400$ rotations of the disk inner
radius for different magnetic diffusivity,
$\eta=0$ ({\it top}),
$\eta=0.01$ ({\it middle}),
and $\eta=0.1$ ({\it bottom}).
{\it Thin} lines denote 30 logarithmically spaced isocontours of density.
{\it Thick} lines denote 20 and twenty linearly spaced magnetic flux
surfaces (or poloidal field lines).
The parameters are 
$\delta_{\rm i} = 100$,
$\kappa_{\rm i} = 100$,
$\mu_{\rm i} = -1.0$,
$\beta_{\rm i} = 0.282$,
and
$v_{\rm inj} = 0.001$.
Note the preserved initial hydrostatic density and force-free field
distribution in front of the bow shocks. 
The figure demonstrates that the bow shock advances slower with increasing
diffusivity.
}
\end{figure*}

As initial condition we choose the same setup as OP97. 
The initial poloidal magnetic field is defined by the current-free potential 
field configuration of the $\phi$-component of the vector potential,
$A_\phi = \left(\sqrt{r^2+(z_{\rm d}+z)^2}-(z_{\rm d}+z)\right)/r$.
The dimensionless disk thickness $z_{\rm d}$ satisfying $(z_{\rm d}+z)>0$ for 
$z<0$ is introduced in order to avoid any kinks in the field distribution.
The initial coronal density distribution is in hydrostatic equilibrium,
$\rho = (r^2+z^2)^{-3/4}$.
The initial corona is defined by two free parameters $\delta_{\rm i}$ and
$\beta_{\rm i}$.

The disk itself as a boundary condition for the jet flow 
is in centrifugal balance and penetrated by a force-free magnetic field.
As the disk boundary condition is time-independent, the initial 
potential field magnetic flux from the disk is concerned.
The toroidal component of the magnetic field in the ghost zones 
($ z<0$) is chosen as $B_\phi(z<0)=\mu_{\rm i}/r$,
where $\mu_{\rm i}$ is another free parameter.
The mass flow rate from the disk surface into the corona is defined by
the injection velocity and the density of the injected material.
With the launching angle $\Theta_0(r,z=0)$ (measured from the jet axis),
the velocity field in the ghost zone is
$ \vec{v}=(v_r, v_{\phi}, v_z) = v_{\rm inj}
( v_{\rm p}\sin\Theta_0,v_{\rm K},v_{\rm p}\cos\Theta_0 ) $ 
for $ r\geq 1$
with $ v_{\rm inj}$ as a free parameter.
For $ r\le 1$ the inflow velocity is set to zero, which actually defines
the inner edge of the disk.
The inflow density is given as $\rho_{\rm d}=\kappa_{\rm i} r^{-3/2}$,
with $\kappa_{\rm i}$ as a free parameter.

Besides the disk ``inflow'' boundary condition, the boundary condition along 
the symmetry axis is set as ``reflecting'', and along the two remaining 
boundaries as ``outflow'' 
(see also Stone \& Norman \cite{stone92a},\cite{stone92b}; OP97; FE00). 
Figure\,1 shows the initial setup of a hydrostatic density distribution together
with the potential magnetic field for the part of the computational box close to
the origin (the region of the ``inner jet'').

Our choice for the free parameters is the following.
We have 
$\delta_{\rm i} = 100$,
$\kappa_{\rm i} = 100$,
$\mu_{\rm i} = -1.0$,
and 
$v_{\rm inj} = 0.001$,
similar to OP97 and FE00 in order to allow for a comparison of the results.
For the plasma-$\beta$ we choose $\beta_{\rm i} = 0.282$ (which is equivalent
to OP97, but is based on the original ZEUS-3D normalization)
or a lower value $\beta_{\rm i} = 0.141$ which has some numerical advantages 
\footnote{See also Appendix A in FE00 for the different choice of 
$\beta_{\rm i}$ in OP97 (equal to unity) and here and in FE00.}.
The lower $\beta_{\rm i}$ does not change the general behavior of the 
jet. 
The jet evolution is faster (in physical time, not in CPU time) and 
the Alfv\'en surface is slightly shifted in $z$-direction,
but the jet internal structure remains very similar.

\subsection{Magnetic diffusion in jets}
Most of the models of MHD jet formation deal with the collimation
and acceleration of a plasma flow in the case of {\em ideal} MHD.
However, it seems to us quite possible that magnetic diffusivity 
may play an important role in protostellar jet formation.
There are (at least) two reasons which may account for that.

The first reason may be the fact that the jet material of young stellar
objects is not fully ionized, in difference from the case of relativistic
jets in AGN or microquasars.
The fraction of ionization (the ratio of ion to neutral particle density)
derived from optical observations is about $0.5 - 0.01$ with the tendency
to decrease along the jet 
(Hartigan et al.\cite{hartigan}, 
 Bacciotti \& Eisl\"offel \cite{bacciotti}). 
From this it can be expected that diffusive effects may indeed affect 
the MHD configuration.
Theoretical studies on this topic have just started recently. 
We refer to Frank et al. (\cite{frank})
who investigated the asymptotic MHD jet 
equilibria under the influence of ambipolar diffusion, showing that
the initial MHD configuration of the jet changes due to ambipolar diffusion
at least on the parsec scale.
Another reference is Ferreira (\cite{ferreira}) who showed the essential role
magnetic diffusion plays for the launching mechanism of the jet from the 
accretion disk.

The other reason for the existence of turbulence in the jet formation region
is the fact that the jet launching object itself -- the accretion disk -- is
highly turbulent.
While turbulence is an intrinsic property of accretion disks
(and {\em necessary} for the accretion process itself), 
turbulence can further be driven in the disk corona by the differential
rotation of the disk, which winds-up the coronal magnetic loops leading to
powerful reconnection processes (see Miller \& Stone 2000).

It seems to be natural to expect that the turbulent pattern being definitely
present in the disk-jet interaction region becomes also advected with the 
jet/wind flow into the domains at greater height above the disk.
We just note that additionally the interaction of the jet with the ambient
medium leading to various kinds of instabilities increase the turbulent 
pattern in the jet (however, we do not expect that the latter process does
affect the jet collimation region).

In Faraday's law (3), the ratio of the first to the second terms in the 
brackets is the magnetic Reynolds number. 
It can be represented by $R_{\rm m}=v\,L/\eta$, 
where $L$ is a typical length scale and $v$ is a typical velocity.
Due to the large length scales in the astrophysical context,
$R_{\rm m}$ based on the microscopic diffusivity is very large
compared to unity.
For a fully ionized hydrogen plasma the microscopic diffusivity is 
$\eta_\mu\sim r_{\rm e} c(v_{\rm th}/c)^{-3}$, 
where $r_{\rm e}=e^2/(m_{\rm e} c^2)$ is the classical electron 
radius,
and $v_{\rm th}=\sqrt{k_{\rm B} T/m_{\rm e}}$ is the electron thermal speed.
For $T=10^4 {\rm K}$, $v=100 {\rm km/s}$, and $L=100$\,AU we obtain 
$R_{\rm m}=v\,L/\eta_\mu\approx 10^{15}$.

It is clear that the relevant diffusivity in the protostellar disk and jet
is most probably ``anomalous'' determined by macroscopic MHD instability 
phenomena (see above), with the resulting magnetic Reynolds number being
much smaller.
This {\em magnetic turbulence} we may parameterize the same way as in a 
Shakura-Sunyaev model for the hydrodynamical viscosity. 
We can define a {\em turbulent} magnetic diffusivity 
$\eta_{\rm T}=\alpha_{\rm m} v L$, where $\alpha_{\rm m}\leq 1$. 
As a characteristic velocity for the dynamical change of the system we might 
choose the poloidal Alfv\'en speed, 
$v_{\rm A}=B_{\rm p}/\sqrt{4\pi\rho}$. 
If we choose the size of our computational box $r_{\rm max}$ as a typical length
scale and $\alpha_{\rm m}\approx 0.1$,
we obtain $R_{\rm m}=v_{\rm A} r_{\rm max}/\eta_{\rm T}\approx 10$ as the
typical magnetic Reynolds number for the {\em global} jet evolution.
In difference, typical length and time scales are different if we are
interested in the evolution of the local jet structure (as needed for
example for the definition of the numerical time stepping in the code).

It might be expected that the diffusivity throughout the jet and the disk 
corona differ considerably.
However, it seems natural to expect the diffusivity in a corona close to 
the disk surface not to differ much from the value in the outer part of the
disk.
For simplicity, and since we are first interested in the general effect, 
our simulations are performed with a constant diffusivity parameter.
To introduce a non-constant diffusivity is straight forward.

\setlength{\unitlength}{1mm}
\begin{figure}
\centering
\includegraphics[bb= 18 144 592 518,width=9cm]{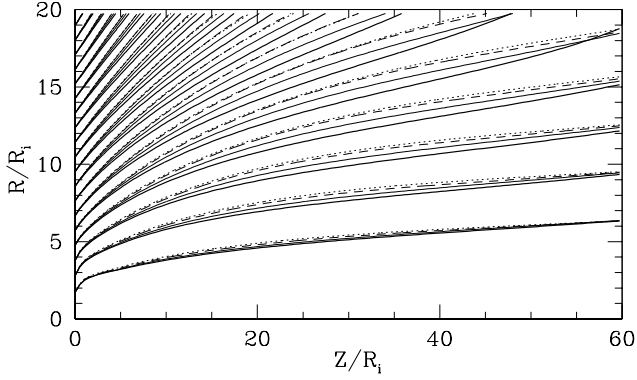}
\caption{
The evolution of the inner jet approaches the quasi-stationary state.
Shown are poloidal magnetic field lines in the case of $\eta = 0.1$
for different time steps, $t = 250, 300, 350, 400$
({\it thick solid, thin solid, dashed} and {\it dotted} lines).
Same parameter setup as in Fig.\,2, except $\beta_{\rm i} = 0.141$.
Grid size $280\times 80$ elements for a physical size 
of $(140\times 40)r_{\rm i}$.
The picture shows how the poloidal magnetic field lines diffuse outwards
but approach a (quasi)-stationary state after 400 rotations 
(see the dashed and dotted lines almost coinciding).
}
\end{figure}

\begin{figure}
\includegraphics[bb= 18 144 592 718,width=4.2cm]{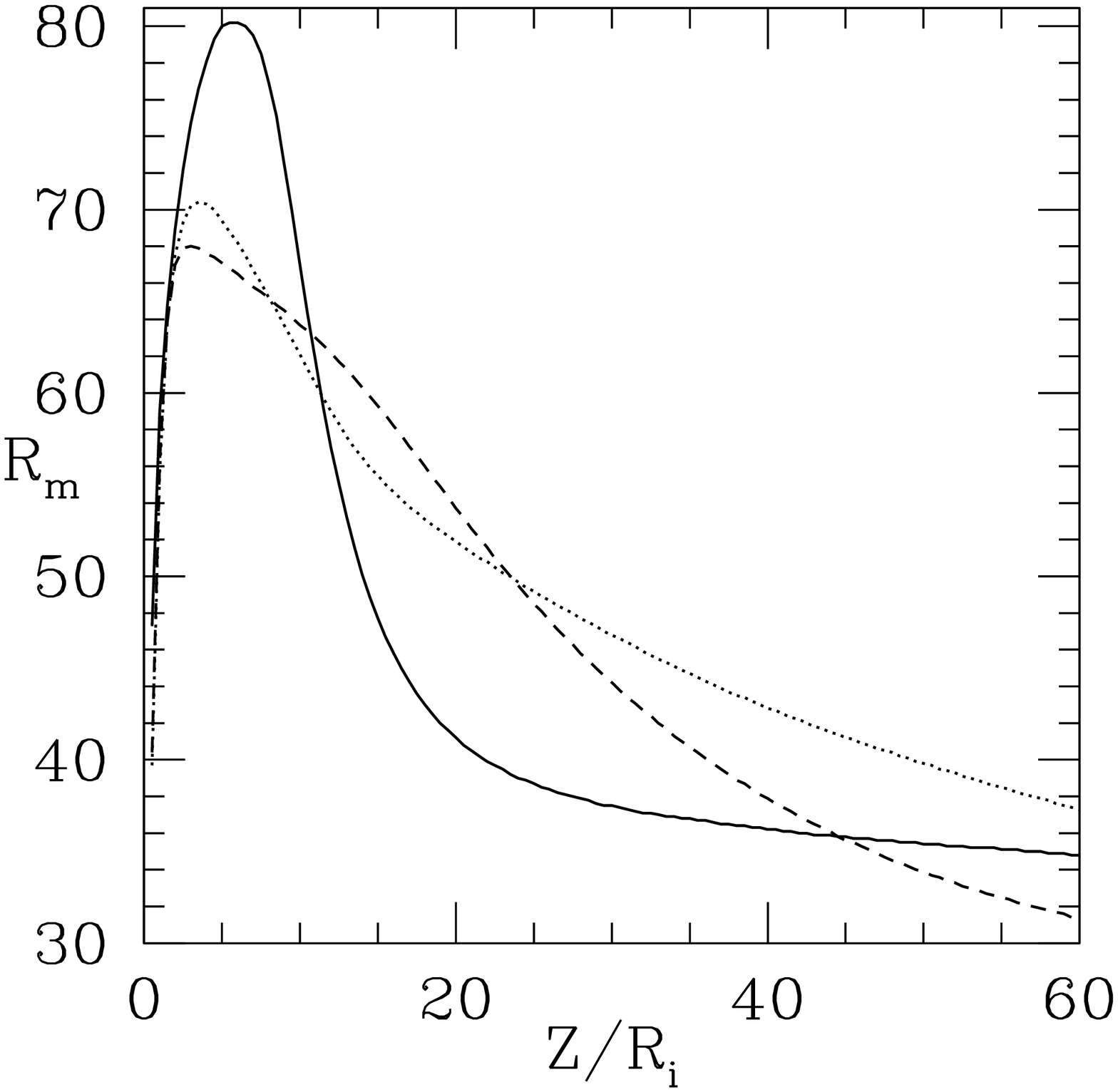}
\includegraphics[bb= 18 144 592 718,width=4.2cm]{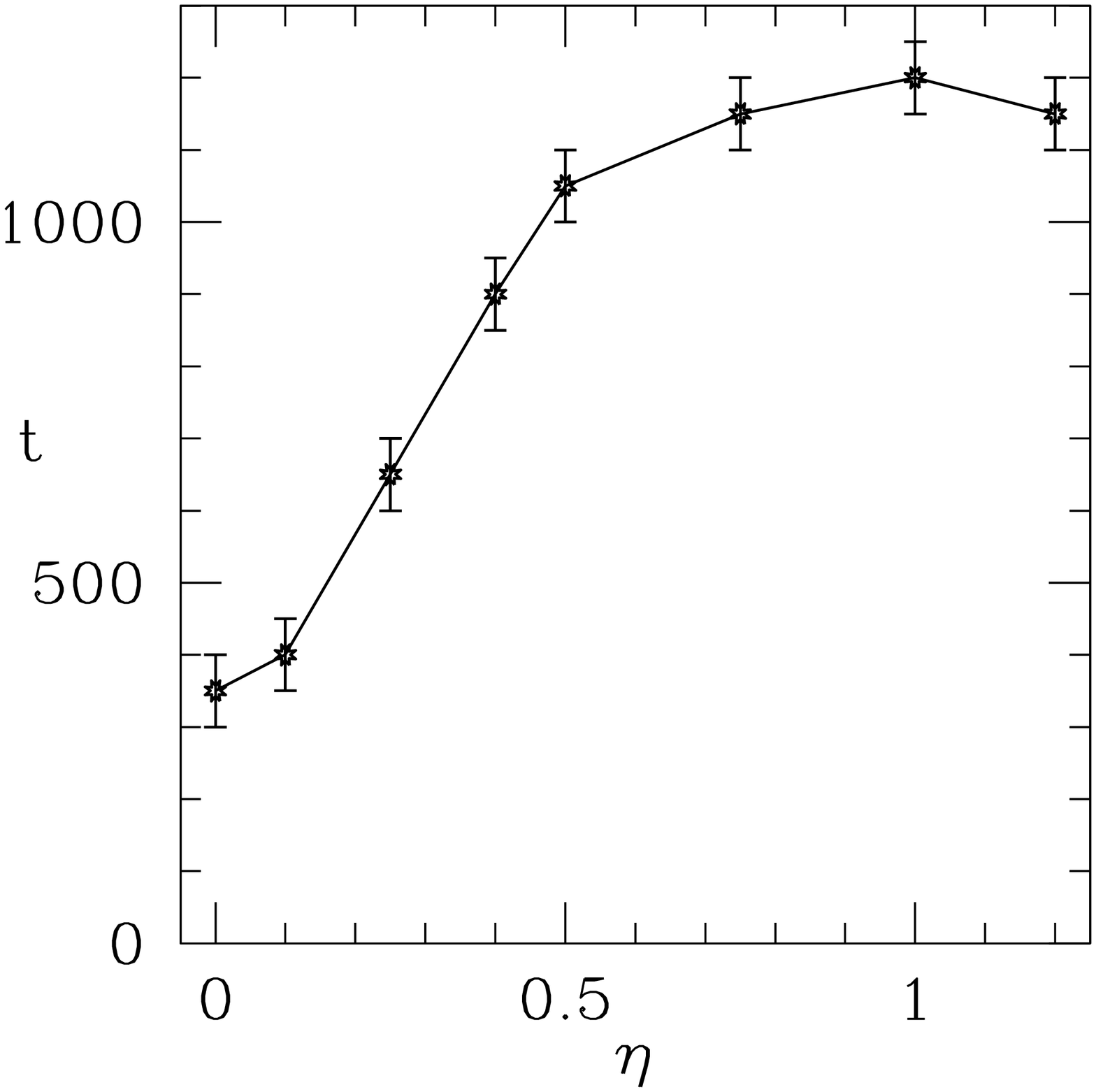}
\caption{Magnetic diffusivity and global time scales.
{\it Left} 
Global magnetic Reynolds number as defined from the scale
of the computational box of the inner jet $L = 20, v = v_{\rm A}$.
Shown is the $ R_{\rm m}$-profile along slices in $z$-direction
at $r =7,13,20$ ({\it solid, dotted, dashed} line, respectively), for
the simulation run with $\eta = 0.1$, $\beta_i = 0.141$.
The plot shows the variation of the typical velocity $v$,
for which we have chosen the local Alfv\'en speed. 
{\it Right} 
Time of stationarity. 
This plot shows for different magnetic diffusivity the time period
when the inner jet reaches a quasi-stationary state.
This time is estimated from the evolution of the poloidal magnetic field
lines (see Fig.\,3) and the error bars indicate our uncertainty. 
}
\end{figure}

The local magnetic Reynolds number is also described by the 
ratio of the dynamic to the diffusive time scale
$ R_{\rm m}=\tau_{\rm diff}/\tau_{\rm dyn}$. 
Here we may define $\tau_{\rm diff}={\rm min}(l^2/\eta)$ and 
$\tau_{\rm dyn}={\rm min}(l/v_{\rm A})$, 
where $l$ is the size of the numerical grid cell. 
Because of the internal structure of the jet,
these minimum values can sometimes be well below the actual {\it global} 
magnetic Reynolds number described above, which is determined by the 
characteristic length scale of our simulation $L=r_{\rm max}$. 

It is interesting to note that in the model of OP97 a {\em turbulent
Alfv\'en pressure}, $p_A$, has been introduced in 
order to establish a pressure equilibrium of a {\em cold} corona 
(or jet) above the (hot) accretion disk.
Although this turbulence effect was taken into account for the pressure
balance, OP97 did not consider it as a reason for a turbulent
{\em magnetic diffusivity}.
However, it is straight forward to relate the Alfv\'enic turbulent 
pressure $p_A$ to a turbulent magnetic diffusivity $\eta_T$.
In order to derive such a relation, 
we now choose the turbulent velocity field $v_{\rm T}$ instead of the local
Alfv\'en speed as the typical velocity for the turbulent diffusivity,
\begin{equation}
\eta_{\rm T} = \alpha_{\rm m} v_{\rm T} L.
\end{equation}
With the definition $\beta_{\rm T} \equiv (c_s / v_{\rm T})^2 $
and taking into account the fact that $c_s^2 = \gamma p/\rho$ for an adiabatic or
polytropic gas law, we obtain
\begin{equation}
v_{\rm T}^2 = \frac{\gamma}{\beta_{\rm T}} \frac{p}{\rho}.
\end{equation}
The normalization gives 
\begin{equation}
{v'}_{\rm T}^2 = 
\frac{p'}{\rho'}\frac{c_{s,i}^2}{v_{K,i}^2}\frac{1}{\beta_{\rm T}},
\quad\quad
{\rm or}
\quad\quad
{v'}_{\rm T}^2 = \rho'^{\gamma-1} \frac{\gamma}{\delta_i\beta_{\rm T}}.
\end{equation}
With the condition of sub-Alfv\'enic turbulence OP97 derived 
$\beta_{\rm T}$ = 0.03.
With a typical value for $\delta_i\simeq 100$ and a ``mean'' value
for the normalized density ${\rho}' \simeq 10^{-2}$
we obtain  for the normalized magnetic diffusivity 
\begin{equation}
\eta' \simeq 0.015\left(\frac{\alpha_{\rm m}}{0.1}\right)
                    \left(\frac{L'}{1.0}\right).
\end{equation}
Since the diffusivity changes only weakly with the density,
$\eta \sim \rho^{1/3}$, this provides a good estimate on the
strength of magnetic diffusion.
A self-consistent simulation would take into account the 
relation between diffusion and density as in Eq.\ (10). 
For comparison,
we run a few of simulations with such an setup.
So far, we find no significant difference to the 
computations with $\eta=const$.

\section{Results and discussion}
We now discuss the results of our numerical simulations
considering the MHD jet formation under the
influence of magnetic diffusion. 
For this, we have run simulations with a different parameter setup.
The simulations were performed 
(i) in domains of different physical size in order to investigate 
the influence of boundaries and to obtain information about the 
large-scale flow, and also 
(ii) with different numerical resolution. 

We detected numerical artifacts (a spurious velocity pattern)
in the corners of the grid where outflow boundary conditions 
meets the other (inflow, reflecting) boundary conditions.
In general, this artifacts remain localized close to the corners of the grid
over many hundred of disk rotations.
All the results discussed in this paper are not affected by these
effects as we mainly concentrate on the inner part of the jet flow.
We performed one reference set of global simulations with high resolution
(numerical mesh of $900\times 200$ grid points,
physical grid of $ (z\times r)=(280\times 40)r_{\rm i}$).
In order to investigate effects which concern only the gross behavior the 
jet flow and not its structure in detail,
we run another set of simulations with lower resolution 
(numerical mesh of $280\times 80$ grid points, 
physical grid of $ (z\times r)=(140\times 40)r_{\rm i}$).
The much faster computation of the low resolution simulations allowed us to
follow the jet evolution for a very long time 
even in the case of a high magnetic diffusivity (up to 4000 disk rotations).
The computational domain was a factor 2 shorter in direction of propagation.
All other parameters were the same as for the high resolution runs.

\subsection{Formation of the global jet}
The first point to be noted when comparing the large-scale evolution of the
diffusive with the non-diffusive flow, is the different speed of the bow shock.
With increasing magnetic diffusivity, the bow shock propagates slower.
The bow shock pattern velocity is about $0.38, 0.35, 0.28$ for 
$\eta = 0, 0.01, 0.1$, respectively (Fig.\,2).
We will later see that, in apparent contrast, the jet velocity increases
with $\eta$. 

As in the case of ideal MHD simulations, before the bow shock front
builds up, 
torsional Alfv\'en waves propagate from the disk surface into the corona
slightly modifying the initial hydrostatic equilibrium.
The super-Alfv\'enic flow catches up with and passes this wave front.

As the bow shock propagates through the corona it leaves behind a cavity
of matter with dilute density and high velocity.
The initially purely poloidal magnetic field becomes more and more helical.
The toroidal magnetic field component is first generated by winding-up
the initial poloidal field due to differential rotation
between the disk and the static corona
but later comes out as a natural result of the MHD flow due to the inertial
forces of the matter.

The internal structure of the jet behind bow shock layer is smoother in the
case of a non-vanishing magnetic diffusion (see Fig.\,2).
We note that also the ``wiggly'' structure in the density distribution close 
to the disk in the case of $\eta=0$ is less prominent in the case of diffusive
simulations.
These ``wiggles'' seem to be a numerical artifact probably due to the density
jump between the disk and the jet, however, the density variations are only
of the order of some percent.

Here, we note another important point.
In this paper we were interested in the cases of a typical MHD jet flow
starting as a sub-Alfv\'enic (but super-slow magnetosonic) flow from the disk
surface, being accelerated to super Alfv\'enic and super-fast magnetosonic
speed, as e.g. described in the paper by Blandford \& Payne (\cite{blandford}).
This case often described as a {\em magneto-centrifugally driven} disk wind/jet
differs from the case where matter is injected into the disk corona already
with super Alfv\'enic speed.
The latter case applies in the case of a relatively weak disk poloidal magnetic 
field as for example in the case of a central dipolar field with a strong gradient
in radial direction.
These winds are initially driven by the (toroidal) magnetic field pressure gradient
in vertical direction
(Lovelace et al. \cite{lovelace}, 
Contopoulos \cite{contopoulos}, 
Fendt \& Elstner \cite{fendt00}).

In this paper we are interested only in the classical case of MHD jet 
formation.
However, in our simulations we note that the Alfv\'en surface moves as a function
of time until the quasi-stationary state (see below) is reached.
For a smaller than moderate magnetic diffusivity ($\eta=0,...,0.5$),
the location of the Alfv\'en surface is always within the active zones well above the
accretion disk boundary.
For higher diffusivity the Alfv\'en surface may advance into the disk for small 
radii and the character of the MHD flow is changed.
We do not consider these cases in our discussion.

\subsection{Inner jet}
In the following we discuss the evolution of the inner substructure of
the global jet close to the jet axis and the accretion disk.
The size of this region is $60\times 20 r_i$ and is comparable e.g. to
the full grid in OP97.
This part of the jet is not influenced from any outflow boundary
condition or the jet evolution on the global scale.
Note that the major part of the global jet is super fast magnetosonic,
hence, there is no (physical) way to transport information
from this part in upstream direction into the inner jet.

\begin{figure}
\includegraphics[bb= 18 144 592 518,width=9cm]{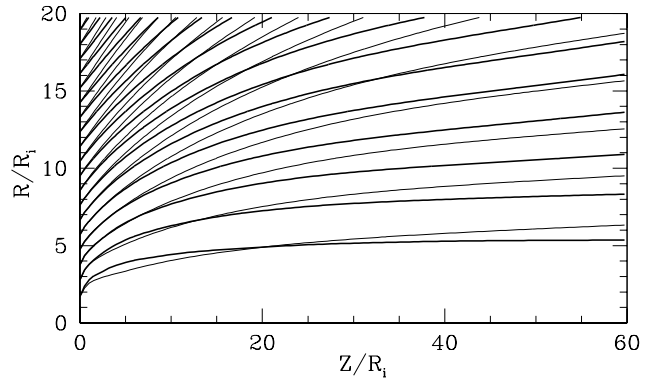}
\caption{The de-collimation of the stationary state poloidal magnetic 
field due to magnetic diffusivity.
Same parameter setup as in Fig.\,3.
Shown is the poloidal magnetic field line distribution of the inner jet
in the state of quasi-stationarity ($t=400$) for vanishing diffusivity 
$\eta=0$ ({\it thick}) and for $\eta=0.1$ ({\it thin}).
}
\end{figure}

%
\begin{figure}
\centering
\includegraphics[bb= 18 144 592 518,width=9cm]{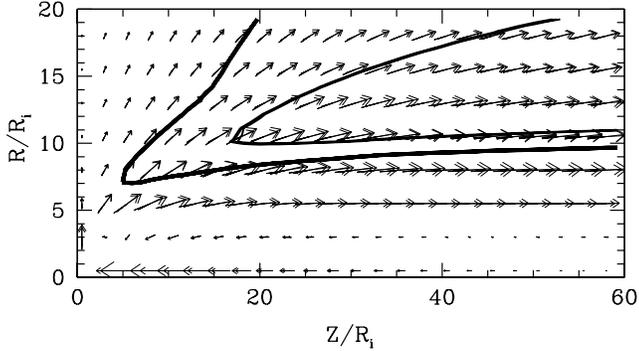}
\caption{ Comparison of poloidal velocity vectors.
Shown are the poloidal velocity vectors at the time of quasi-stationarity.
Over-plot of the velocity field for the $\eta = 0$ simulation
at $t=350$ with the $\eta = 0.1$ simulation at $t=400$.
The de-collimation visible in the poloidal magnetic field lines
(see Fig.\,5) is {\em not} present here. 
The thick line indicates the Alfv\'en surface for $\eta = 0.1$,
the thin line the fast magnetosonic surface.
}
\end{figure}

\subsection{Quasi-stationary nature of the (inner) jet}
As a major outcome of OP97 and FE00 it has been found that under a 
certain choice of boundary conditions the disk outflow may evolve 
into a stationary state after several hundreds of disk rotations.
Such a stationary state solution provides an ideal test bed for
investigation of the internal forces acting in the jet, since
the well known equations of axisymmetric ideal MHD and its 
conservation laws may be used for interpretation of the jet flow.

In our present simulations considering the MHD jet formation under the 
influence of magnetic diffusion, we also find such stationary states.
We denote them as {\em quasi-stationary} since -- due to the 
large computational domain -- such a behavior can be seen only
in the inner region within reasonable computational time.
The outer regions surrounding the (inner) stationary jet flow 
will further evolve in time.
For the area of the inner jet the stationary state is reached 
after approximately $t=350$ disk rotations in the case of
ideal MHD.
This  time scale can be quite different in the case of a
non-vanishing magnetic diffusion and also depends on the
plasma-beta parameter $\beta_{\rm i}$.
Note that in spite of considering magnetic diffusivity, we use
the same parameter setup as OP97 and FE00.

Figure 3 shows an over-plot of poloidal magnetic field lines 
resulting from our simulations at the time steps of
$t=250, 300, 350, 400$ disk rotations, respectively,
in the case of relatively large magnetic diffusion $\eta=0.1$.
It can clearly be seen how the poloidal magnetic field lines first 
diffuse outwards from their position at $t=250$ (which is
close to the non-diffusive field distribution, see also Fig.\,5).
After some hundred of rotations more, the field distribution
approaches the quasi-stationary state
(note the dotted lines almost matching the dashed ones). 

First, we note that the existence of such a quasi-stationary state
might not necessarily be expected in the case of magnetic diffusion
and instead one might think that the magnetic field will just decay
forever.
The reason, why a stationary state is possible in the case of a jet
flow, is that with such a setup a loss of magnetic energy in the 
jet caused by diffusion can be replenished by the constant
Poynting flux rising from the disk.
This energy reservoir can be thought of as established
constantly by the disk rotation and accretion and eventually by the
gravitational potential of the central star.
Note that also in the ideal MHD case the jet flow carries energy out of the
computational box. Also this energy loss is balanced by the same way.
The additional effect due to magnetic diffusivity is small compared to the
total energy flow in the jet.

In general, our simulations show that with increasing $\eta$
the flow reaches the quasi-stationary state at a {\em later} time.
For a smaller than moderate magnetic diffusivity ($\eta=0,...,0.5$),
we find an approximately linear relation between this time and the
diffusivity (Fig.\,4, right panel). 
This is also the range where the jet flow is a classical MHD jet
like the Blandford-Payne solution, starting as a sub-Alfv\'enic flow
from the disk surface and crossing the Alfv\'en surface at some height
above the disk.
For higher diffusivity the Alfv\'en surface has been advanced into
the disk for small radii and the character of the MHD flow is changed.
We do not consider these cases in particular, however, we show the
time of quasi-stationary state for completeness.
 
For comparison, 
the local time step accounting for magnetic diffusion and the Alfv\'en
time step, 
together with the {\it global} magnetic Reynolds number $R_{\rm m}$ 
(defined by the jet size and a mean value for the jet Alfv\'en speed,
see Sect.\,3.3) 
and the related $\alpha_{\rm m}$ are given in the Appendix in Tab.\,B.1 for 
the time when
the quasi-stationary state is reached.
In Table~B.1 we have chosen a ``typical'' value for the Alfv\'en speed 
within the grid of the inner jet.
That this is feasible, is shown in Fig.\,4 ({\it left}) 
where we plot the variation
of $R_{\rm m}$ along the jet in the case $\eta=0.1$.
As $R_{\rm m}$ does not change along the jet by more than a factor of two
(this is similar for other diffusivity), this value provides a good
estimate for the global jet dynamical behavior.

\begin{figure*}

\includegraphics[bb= 18 244 592 718,width=8cm]{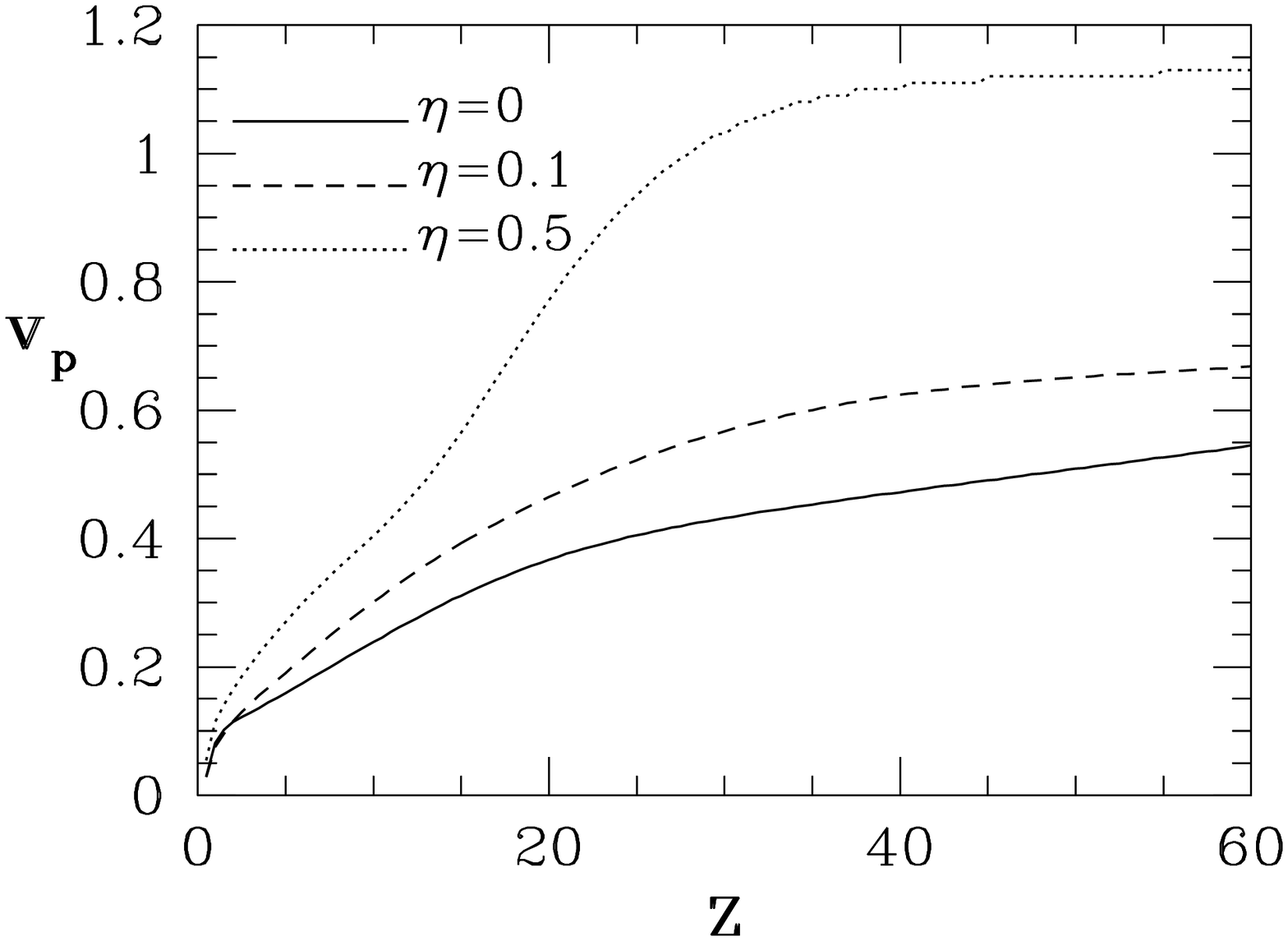}
\includegraphics[bb= 18 244 592 718,width=8cm]{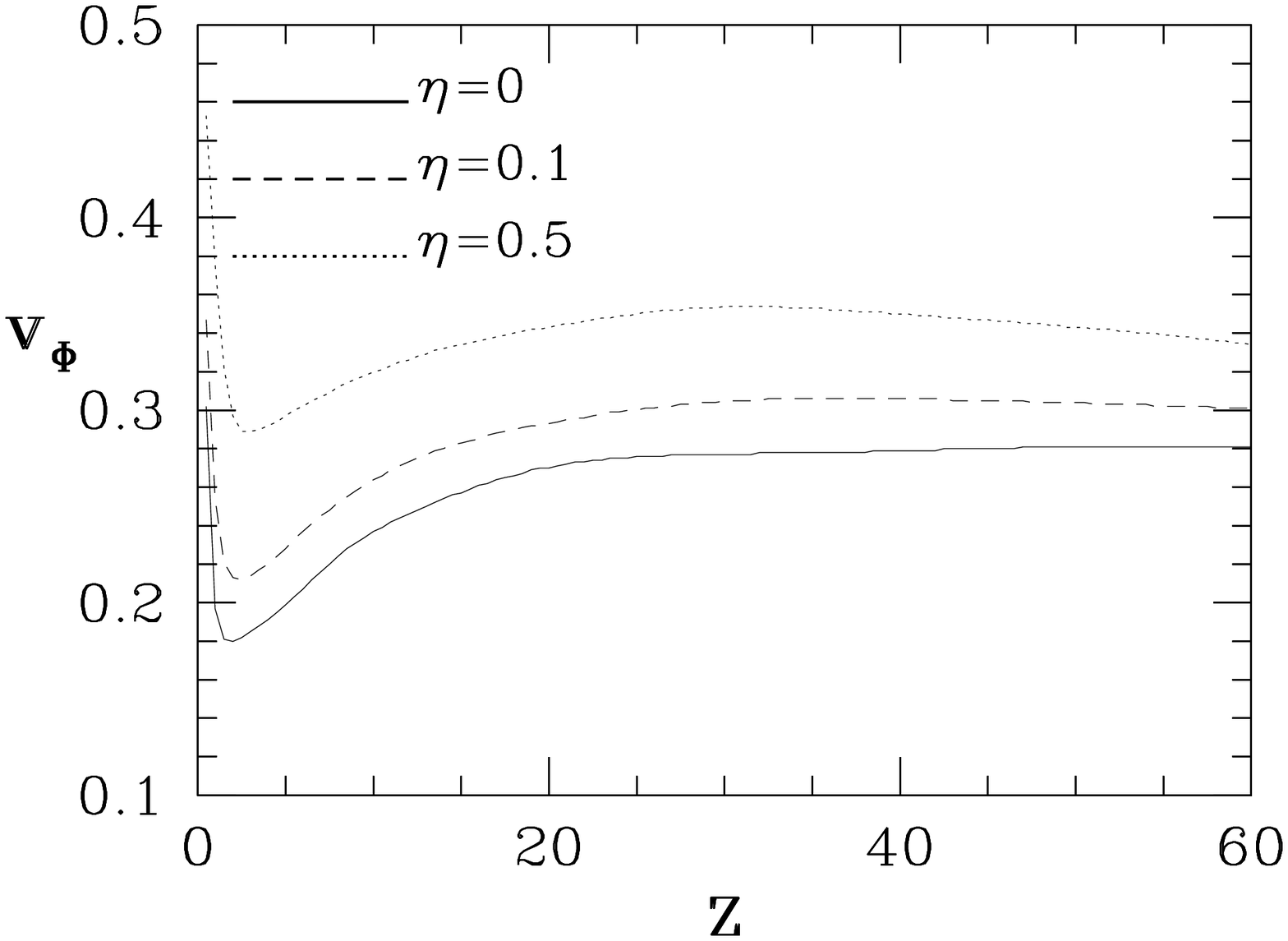}

\includegraphics[bb= 18 244 592 718,width=8cm]{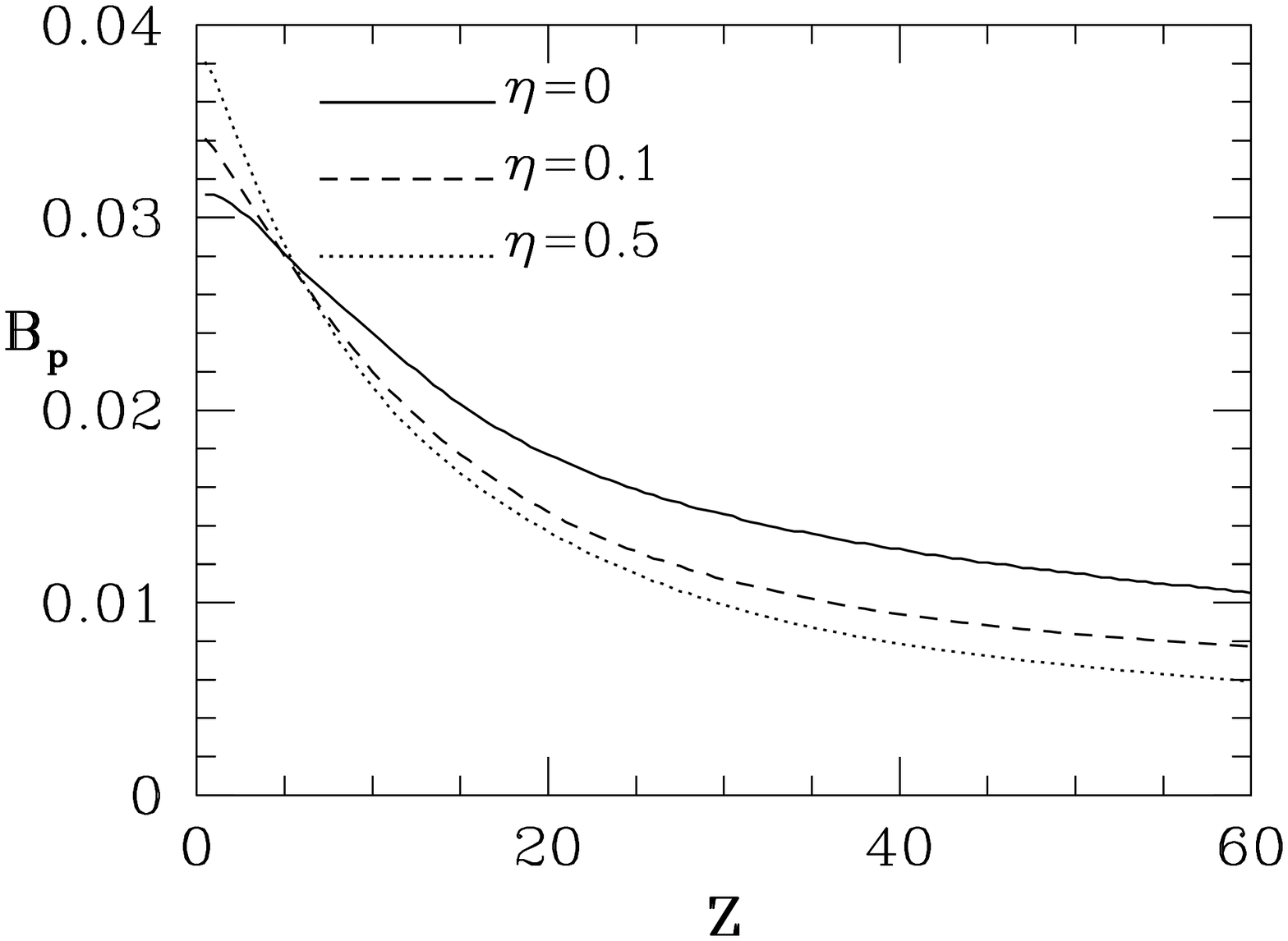}
\includegraphics[bb= 18 244 592 718,width=8cm]{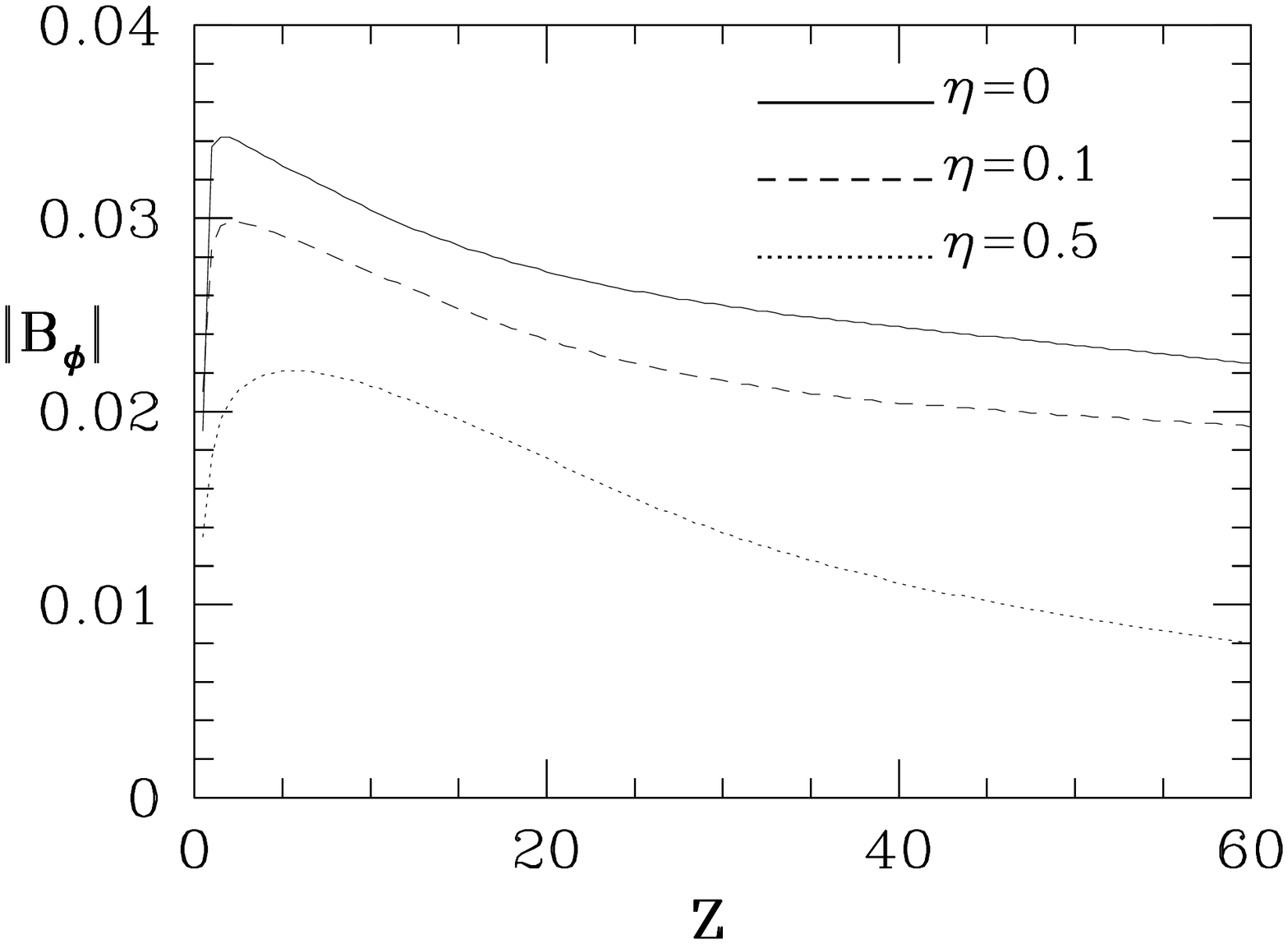}
\caption{Variation of the jet velocity and magnetic field with
different magnetic diffusivity, 
$\eta=0$ ({\it solid}),
$\eta=0.1$ ({\it dashed}),
$\eta=0.5$ ({\it dotted}).
Shown is the profile of the poloidal ({\it left}) and toroidal 
({\it right}) components 
of the velocity ({\it top}) and the magnetic field ({\it bottom})
in $z$-direction at $r=15$ at the time of
quasi-stationarity.
The velocity components increase, while
the magnetic field strength decreases with increasing $\eta$. 
Note the different scales for $v_p$ and $v_{\phi}$.
The boundary value for the toroidal velocity is the Keplerian
value $v_{\phi} = 0.258$ at $r=15$.
}
\end{figure*}

\subsection{Jet velocity and collimation}
The most interesting, since directly observable, quantities of a jet 
are its velocity and degree of collimation.
In Figure 6 we show the poloidal velocity vectors in the inner jet
for a simulation without and with considering magnetic diffusion 
($\eta = 0$ and $\eta = 0.1$), 
taken at that time step, when the flow has reached the quasi-stationary
state.

The general point to mention is that the figures clearly show the 
self-collimating property of the MHD flow as the velocity vectors
become more and more aligned with the jet axis as we go along the flow.
However, there exists also a region of low collimation close to the
disk where the velocity vectors point in radial direction 
(with $\simeq 45\degr$ half opening angle).

For the low magnetic diffusivity ($\eta<0.05$) simulations we have
observed an interesting feature.
The apparent de-collimation of the poloidal magnetic field
structure (Fig.\,5), which is already present for a weak magnetic
diffusivity ($\eta = 0.01$), 
is, however, not visible in the poloidal velocity (Fig.\,6). 
In both cases, the flow evolution has reached a quasi-stationary
state ($t=400$).
In contrast to ideal MHD, in the case of diffusive MHD,
a miss-alignment between $v_p$ and $B_p$ is possible. 
Up to $\eta = 0.5$ 
the mismatch between the poloidal velocity and magnetic field
vector is relatively weak for large $z$ and about $5\degr - 10\degr$
at intermediate heights above the disk.
That means that we generally get a collimated stream along the axis,
however, the mass load distribution varies implying a variation of the mass
flow rate through the $r$ and $z$-boundaries with $\eta$.
If a larger magnetic diffusivity ($\eta>0.1$) is applied, we 
observe also a de-collimation of the {\em mass flow} (see below).

In general, both the poloidal and toroidal velocity increase with 
increasing diffusivity.
At the same time, the magnetic field components decrease with
the increasing diffusivity.
This is shown in Fig.\,7 where we plot the velocity and field
components along jet at a distance of $15\,r_i$ from the jet axis.
Thus, as a conclusion, the diffusive jet becomes faster.
We discuss that point below in the context of the Lorentz forces
acting in the jet.

However, two points should be mentioned concerning Fig.\,7.
The first is the {\em decrease} of toroidal velocity just above the
disk surface. 
As the field line foot points rotate with Keplerian speed and 
the matter is corotating with the field lines,
one would actually expect an increase of toroidal velocity,
if the magnetic field is dominated by the poloidal component.
In our case, we find at a certain radius along the disk surface
that $B_{\phi} = B_{\rm p}$.
Thus, the matter may substantially {\em slide along the field}.
Just above the disk, the toroidal field strength first 
increases with height\footnote{This feature can be also observed
in the simulations of OP97 (see their Fig.\,4)},
and the slide along the field line becomes larger.

The second point is the fact that -- unlike the poloidal velocity --
the magnetic field strength and the toroidal velocity do not seem
to match the given boundary condition along the disk surface for 
the case of non-vanishing diffusivity.
The reason is the jump in diffusivity between the disk (boundary
condition) and the disk corona (active zones of the grid). 
The magnetic field lines are frozen-in the accretion disk, but when
leaving the disk surface, they are immediately affected by 
magnetic diffusion.
Thus, the magnetic field strength in the active zones of the grid
(which are shown in Fig.\,7) deviates from the boundary value
in the case of a non vanishing diffusivity.
In the stationary state solutions shown in Fig.\,7 an equilibrium 
state has been reached between magnetic field diffusion and 
advection.
We see that at this radius ($r=15$) the field strength has 
increased for the region immediately above the disk.

The observed de-collimation of the matter flow with increasing
diffusivity is most evident if we plot the mass and momentum
fluxes across the boundaries of the inner jet region.
We define the fluxes across surfaces parallel to the 
accretion disk boundary by
\begin{equation}
\dot{M}=\int_{0}^{r_{\rm max}}2\pi r \rho v_z dr\ ,
\quad \dot{M} v_z = \int_{0}^{r_{\rm max}}2\pi r \rho v^2_z dr\ .
\end{equation}
These are the mass flux and the kinetic $z$-momentum 
(i.e. momentum in $z$-direction) flux along the jet axis. 
For the inner jet $r_{\rm max}=20$ and the integration is
along $z_{\rm max}=60$.
The flux away from the jet axis 
(thus, in r-direction\footnote{Note that we define 
the momentum flux in $z$-direction across the $r_{\rm max}$-boundary
as $\dot{M} v_z = \int_{0}^{r_{\rm max}}2\pi r \rho v_z v_r dr$}) 
is defined correspondingly by the integration along the 
$r_{\rm max}$-boundary from $z=0$ to $z=z_{\rm max}$.
The corresponding flux into the jet 
(thus, prescribed by the disk boundary condition)
has to be integrated along the $z=0$ axis.
Signature of a good degree of collimation would be the fact that most of
the mass and/or momentum flux is directed along the jet axis.

Figure 8 shows how the mass and momentum fluxes for different diffusivity
evolve in time.
We show the mass flux across the $r$ and the $z$-{\em boundary} and the
kinetic momentum flux in $r$ and the $z$-{\em direction} integrated along 
both outflow boundaries.

The large mass and momentum fluxes for the outflow during the first 
$100 - 200$ rotations result from the fact that at these stages the
initial hydrostatic corona is being pushed out of the grid of the 
inner jet.
After the the bow-shock has left the inner grid, this initial coronal 
mass reservoir has been swept out and the remaining mass flow in the
jet is given purely by the mass injection rate from the disk boundary 
condition.

In the stationary state the mass inflow from the disk boundary into the jet
must be equal to the mass loss across the boundaries in
$r$- and $z$-direction.
That this is true in our simulations
can be seen in Fig.\,8 on the long time scale if we compare the solid line
(inflow condition) to the sum of the dotted (radial outflow) and the dashed
lines (axial outflow).
The analytical value for the mass rate as integrated from the given
inflow boundary condition is $\dot{M} = 1.41$ which is in good agreement
with the numerical result.
Note that for the momentum flux a similar analytical integration gives a
momentum flux from the disk into the jet of
$\dot{M}v_z = 1.8\times10^{-4}$, 
which is much below the numerical value at the first active zone.
As the momentum flux is not conserved as the matter becomes accelerated in 
the jet, 
this shows the tremendous gain of kinetic energy of the MHD flow.
As a good estimate, the kinetic momentum flux in $z$-direction across the
$z=z_{\rm max}$ boundary is just the integrated mass flow rate
$\dot{M}$ times the mean $v_z$-velocity at this position.
In the case of vanishing diffusivity we have 
$\dot{M}v_z \simeq 1.5 \times 0.6 = 0.9$ which is similar to our numerically
integrated momentum flux\footnote{Almost no kinetic momentum flux in 
$r$-direction leaves the inner box across the $z$-outflow boundary}.

Now we compare the fluxes for simulations with different magnetic
diffusivity.
For the simulation run with $\eta=0.5$ (Fig.\,8, bottom panel)
we have a mass inflow rate of 1.5 (in dimensionless units). 
The mass loss rate across the grid boundaries 
is about 0.45 in $z$-direction and 1.05 in $r$-direction.
Compared to the corresponding values in the case of $\eta=0$, 
where about 70\% of the mass flow leaves the box in $z$-direction, 
this clearly shows that the mass flow for $\eta=0.5$ is less collimated.
This situation is even more evident for the simulations with higher 
diffusivity $\eta=1.0$, where, however, the mass injection
from the disk boundary is partly super-Alfv\'enic (not shown).
Thus, even if the velocity vectors have more or less the same direction
for diffusivity up to $\eta=0.5$, the mass load along the stream lines
is different due to the fact the matter, driven by centrifugal forces,
may diffuse outwards across the magnetic field lines enhancing the
mass flow rate in radial direction.

At this point it might be interesting to discuss the results of recent
diffusive MHD simulations of the jet formation out of the
accretion disk (Kuwabara et al.~\cite{kuwabara}).
These authors find that the jet launching from the disk critically
depends on the strength of the magnetic diffusivity.
For small diffusivity, mass accretion in the disk and jet formation
take place occasionally. 
For intermediate diffusivity the disk-jet system may reach a steady
state.
For high diffusivity the accretion rate and outflow rate decrease with
diffusivity and may even vanish.
In respect to our results, these results imply that the only way to
launch a stationary MHD jet is indeed to allow for a reasonable amount
of magnetic diffusion.
Further, as we find less collimation for higher diffusivity or,
equivalently, a weaker {\em jet}, 
such a state of stationary jet formation may become less important 
as the mass flow rates in the disk and the outflow decrease
substantially.
It would therefore be of great importance to follow the simulations
of Kuwabara et al.~(\cite{kuwabara}) for longer time scales comparable
to our runs.

If we eventually define the {\em degree of jet collimation} by the
{\em mass flux} across the jet boundaries,
our simulations reveal the existence of a {\em critical value} of the
magnetic diffusivity in this respect.
In Fig.\,9 the ratio of the mass flux leaving the grid in $z$-direction
to that in $r$-direction is shown for different magnetic diffusivity for the
time when the bow shock has left the inner box.
This figure shows directly that for a high diffusivity mass flux ratio 
exceeds unity, indicating a weakly collimated mass flow.
Figures 8 and 9 clearly show that for our model setup there exists
a {\em critical value of the magnetic diffusivity}, $\eta_{\rm cr}$.
In the simulations with $\eta \leq \eta_{\rm cr}$ the MHD flow
evolves into a collimated stream.
In contrary, for $\eta \geq \eta_{\rm cr}$ the flow remains
only weakly collimated.
The actual value for the critical $\eta$ depends on the plasma beta $\beta_i$.
In our standard setup chosen for Fig.\,8 and 9, we find
$\eta_{\rm cr} \simeq 0.3$,

We note that the momentum flux gives somewhat different picture. 
The momentum flux in $z$-direction is always larger than that in 
$r$-direction.
For our setup we obtain a ratio of about 5 -- 8 when we compare the
momentum fluxes in each direction.
This demonstrates first the very high efficiency of rotating MHD flow
in converting rotational kinetic energy into poloidal kinetic flux.
%
%
%
In this respect, if we would define the degree of collimation by the
momentum fluxes, our jets would perfectly collimated also for higher
diffusivity.
This leaves the question of how the degree of jet collimation is
properly defined.
Clearly, for diffusive MHD jets the field structure is {\em not}
an accurate measure of propagation.
What concerns the observational appearance, the mass flow
distribution (or actually the density distribution) would be
the theoretical equivalent to the observed intensity (as long
as no emission maps can be provided by the simulations).

In summary, 
we propose that the mass flux gives the best measure of the degree
of collimation.
In our simulations we see a strong indication for the existence of a
critical value of the magnetic diffusivity beyond which such a collimation
cannot really be obtained.

%
\begin{figure*}

\includegraphics[bb= 18 244 592 718,width=8cm]{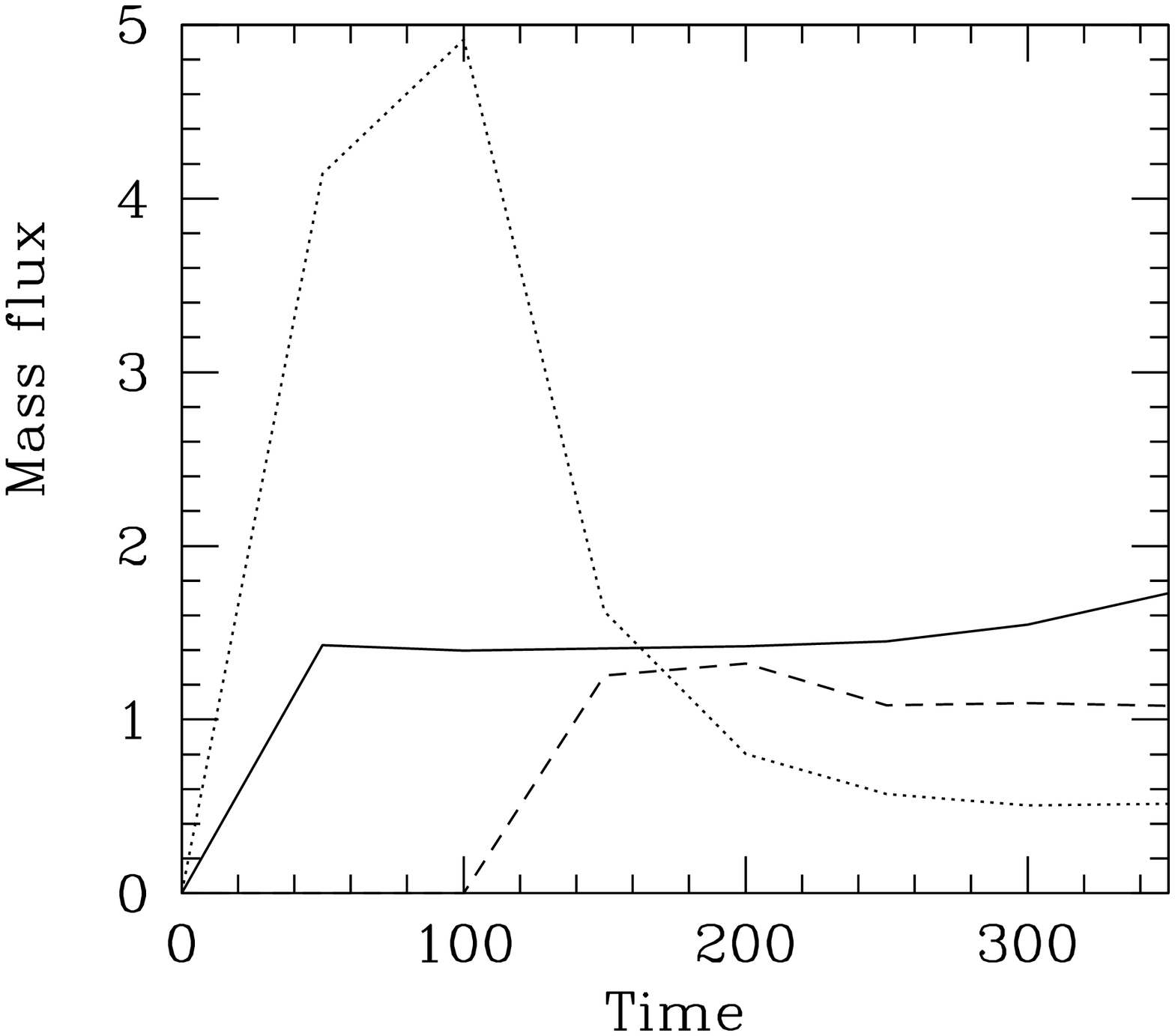}
\includegraphics[bb= 18 244 592 718,width=8cm]{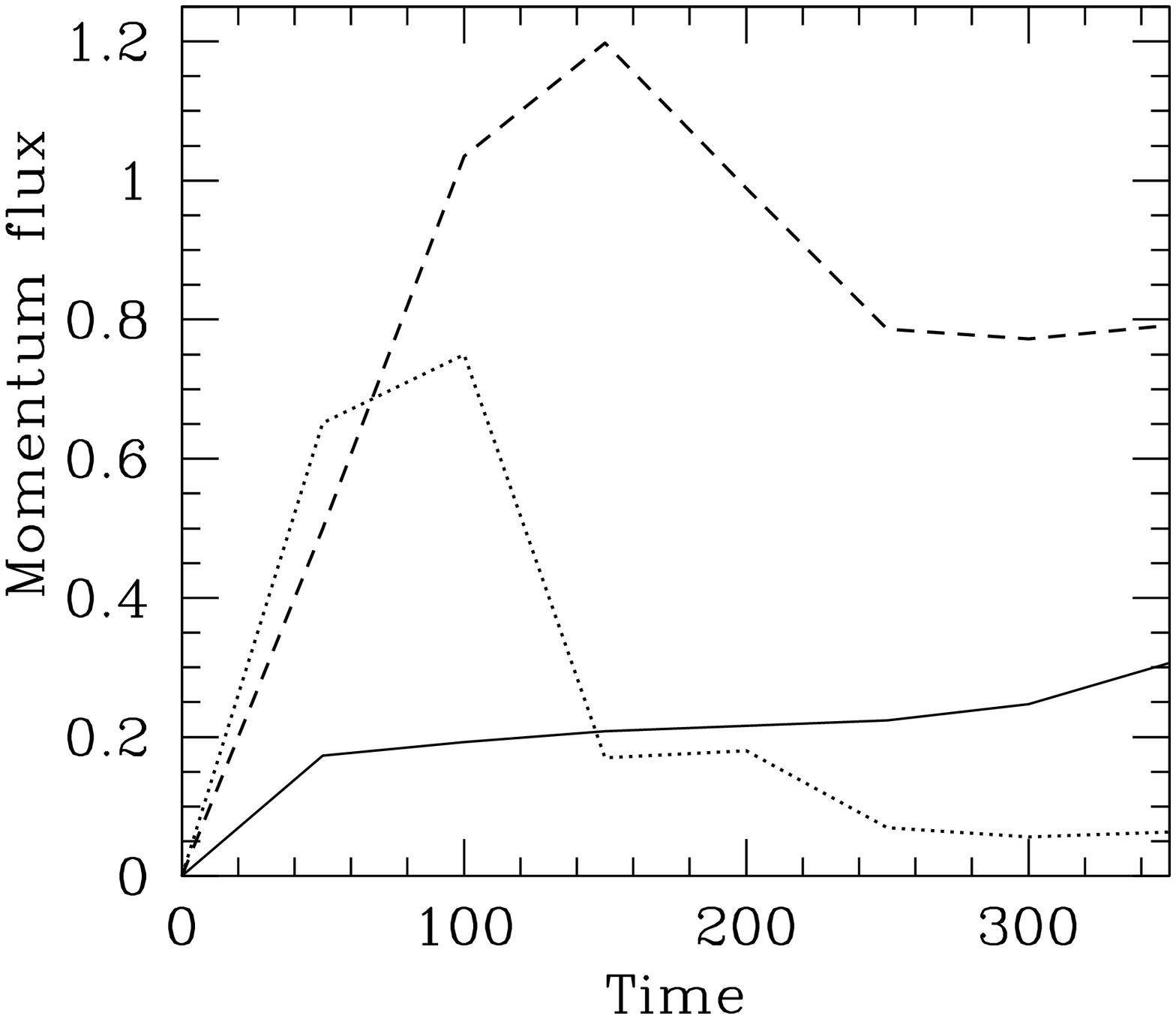}

\includegraphics[bb= 18 244 592 718,width=8cm]{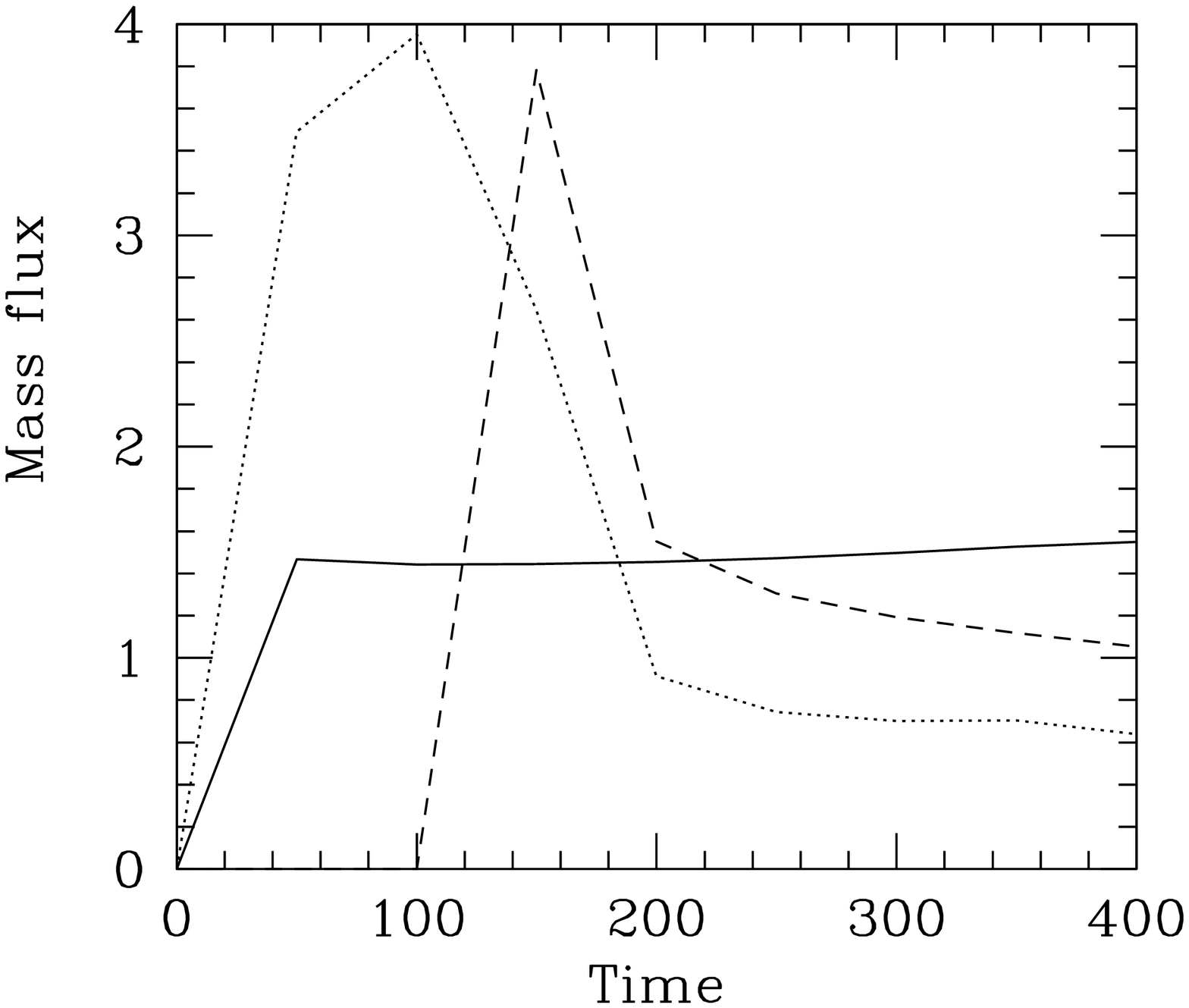}
\includegraphics[bb= 18 244 592 718,width=8cm]{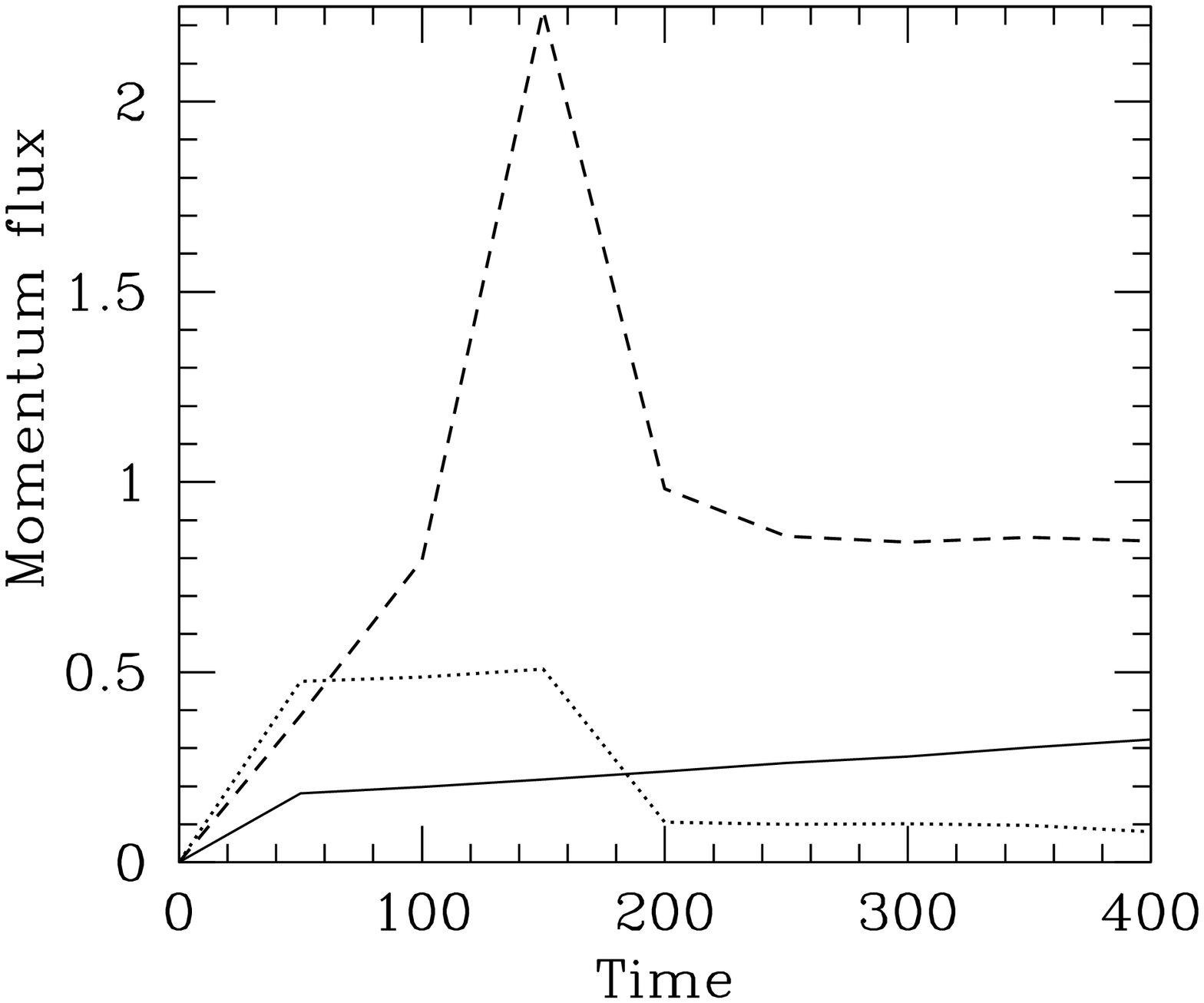}

\includegraphics[bb= 18 144 592 718,width=8cm]{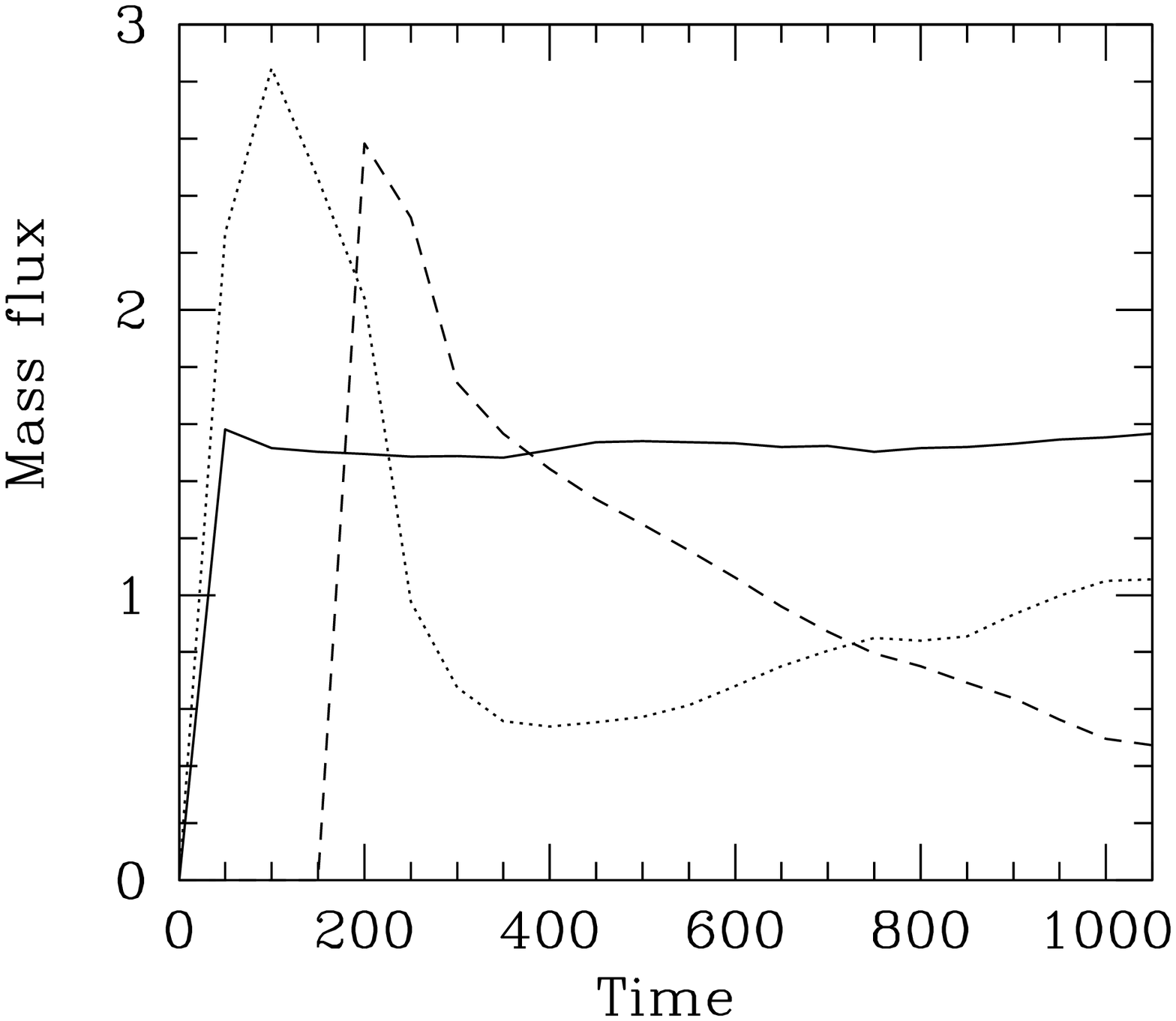}
\includegraphics[bb= 18 144 592 718,width=8cm]{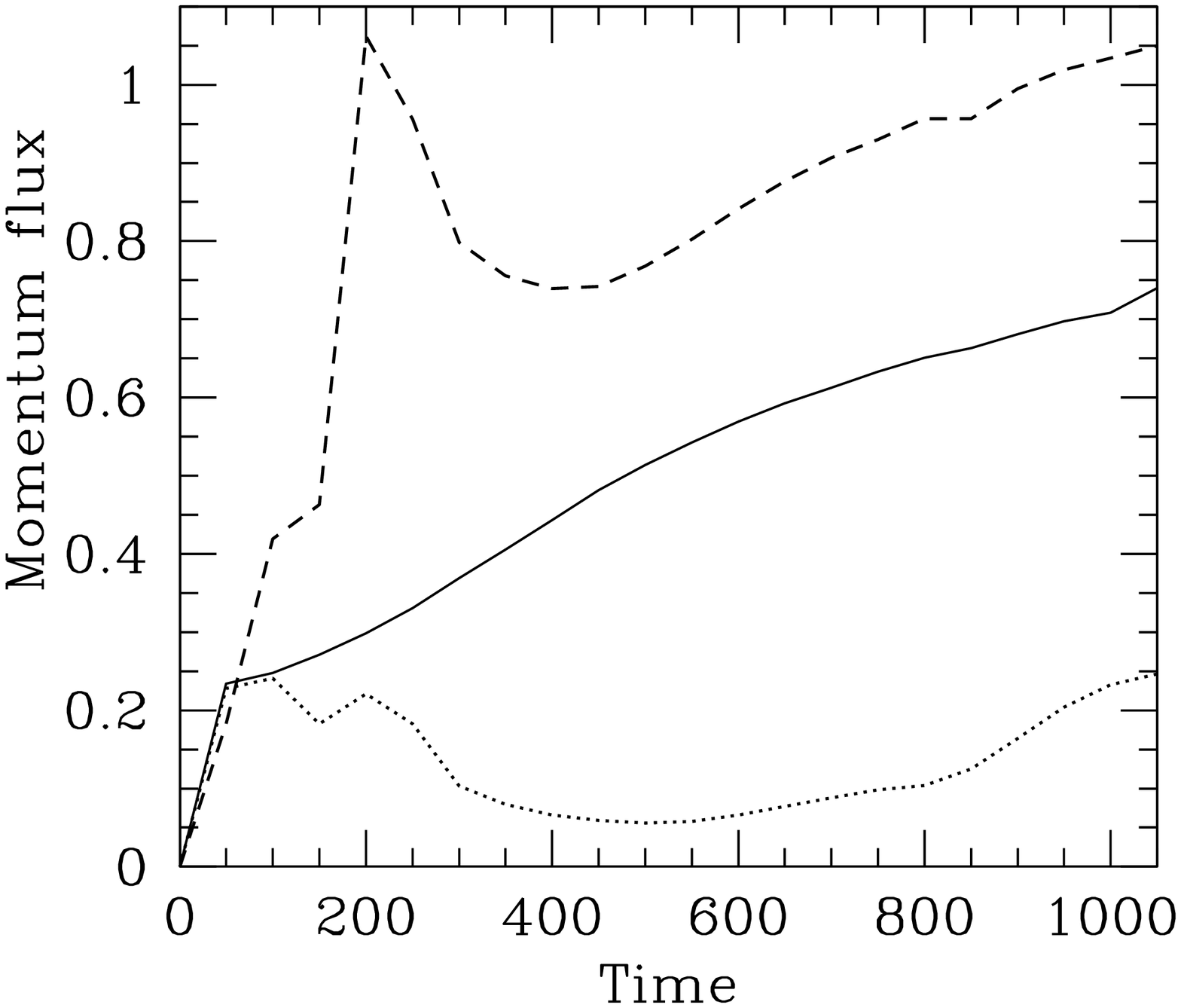}
\caption{
Time evolution of the mass flux and kinetic momentum flux 
for different magnetic diffusivity, $\eta=0,0.1,0.5$ 
({\it top}, {\it middle} and {\it bottom} figures, respectively),
in the inner part of the jet,
$(z\times r)=(60\times 20) r_{\rm i}$.
The final point of each line corresponds to the end of the simulation
when the (quasi-) stationary state has been reached.
Shown is the mass flux ({\it left}) across the different boundaries.
The mass inflow across the first active grid cells along the ($z=0$)-boundary 
({\it solid}),
across the outer ($ z=z_{\rm max}$) axial boundary
({\it dashed}),
and  across the outer ($ r=r_{\rm max}$) radial boundary
({\it dotted}).
Also shown is the kinetic momentum flux across the boundaries ({\it right}).
Poloidal kinetic momentum flux across the first active grid cells
along the ($z=0$)-boundary ({\it solid}). 
Note that this is already evolved from the value of the boundary condition.
Momentum flux in $z$-direction integrated along the outflow boundaries
({\it dashed}).
Momentum flux in $r$-direction integrated along the outflow boundaries
({\it dotted}).
}
\end{figure*}

\begin{figure}
\centering
\includegraphics[bb= 18 144 592 718,width=9cm]{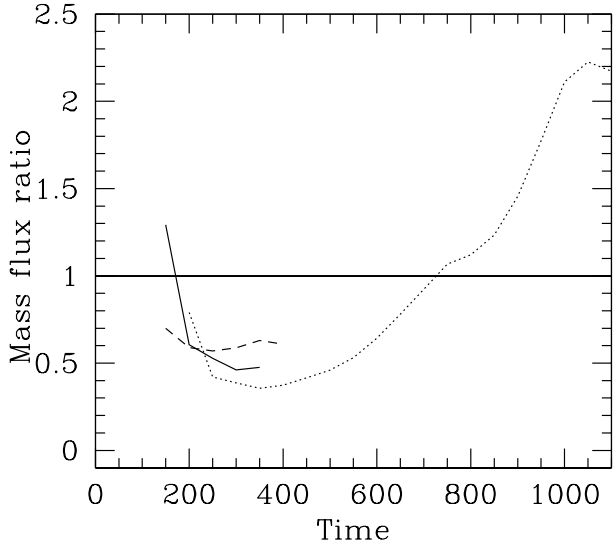}
\caption{
Time evolution of the mass flow ratio between 
the radial outflow boundary 
(mass flow in $r$-direction)
and the axial outflow boundary 
(mass flow in $z$-direction) 
for different magnetic diffusivity, $\eta=0,0.1,0.5$ 
({\it solid}, {\it dashed} and {\it dotted} line, 
respectively),
in the inner part of the jet,
$(z\times r)=(60\times 20) r_{\rm i}$.
The final point of each line correspond to the end of the simulation
when the (quasi-) stationary state has been reached.
For higher diffusivity, the mass flux ratio in the quasi-stationary
state increases indicating a decrease in degree of collimation.
}
\end{figure}
%

\begin{figure*}

\includegraphics[bb= 18 144 592 718,width=7cm]{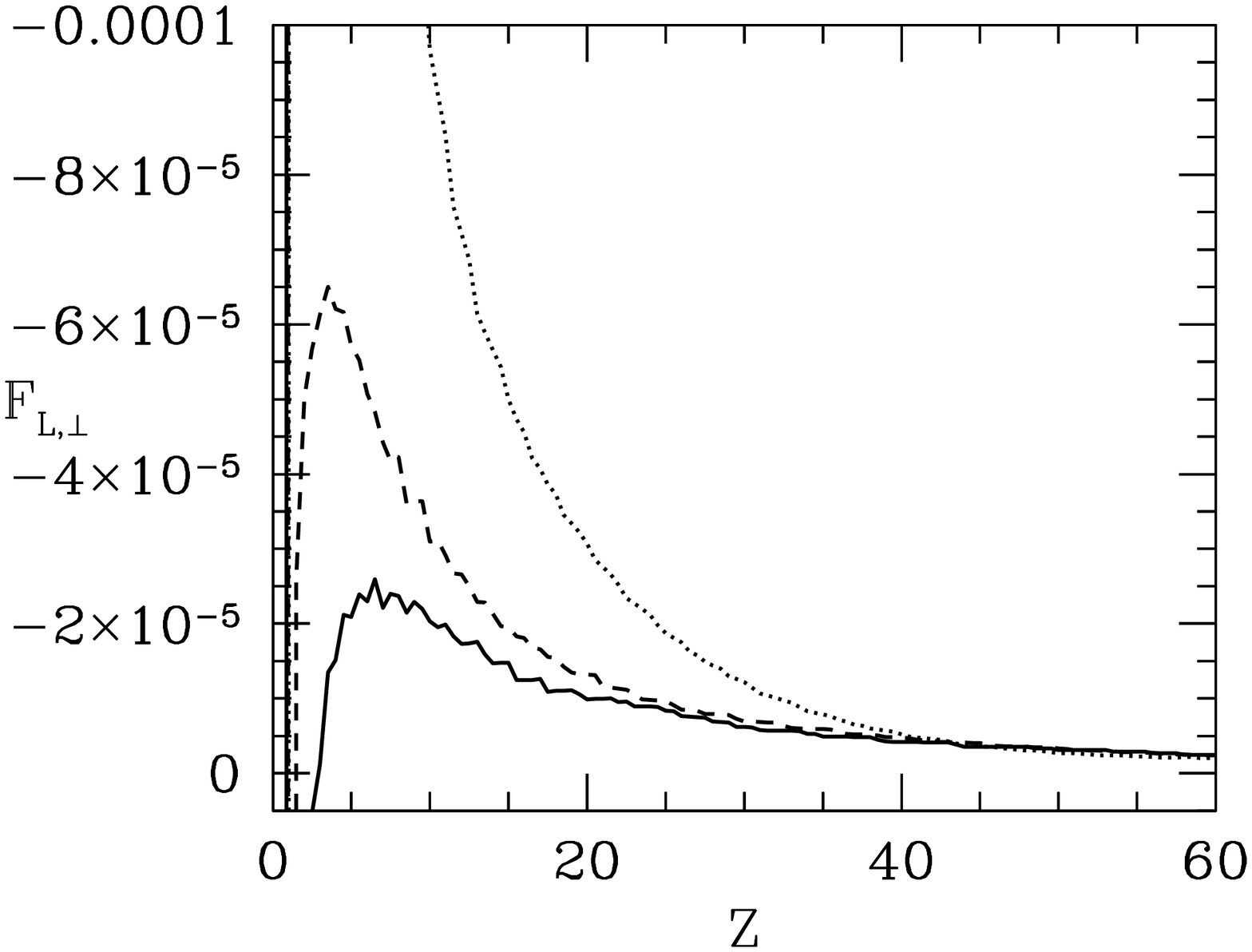}
\includegraphics[bb= 18 144 592 718,width=7cm]{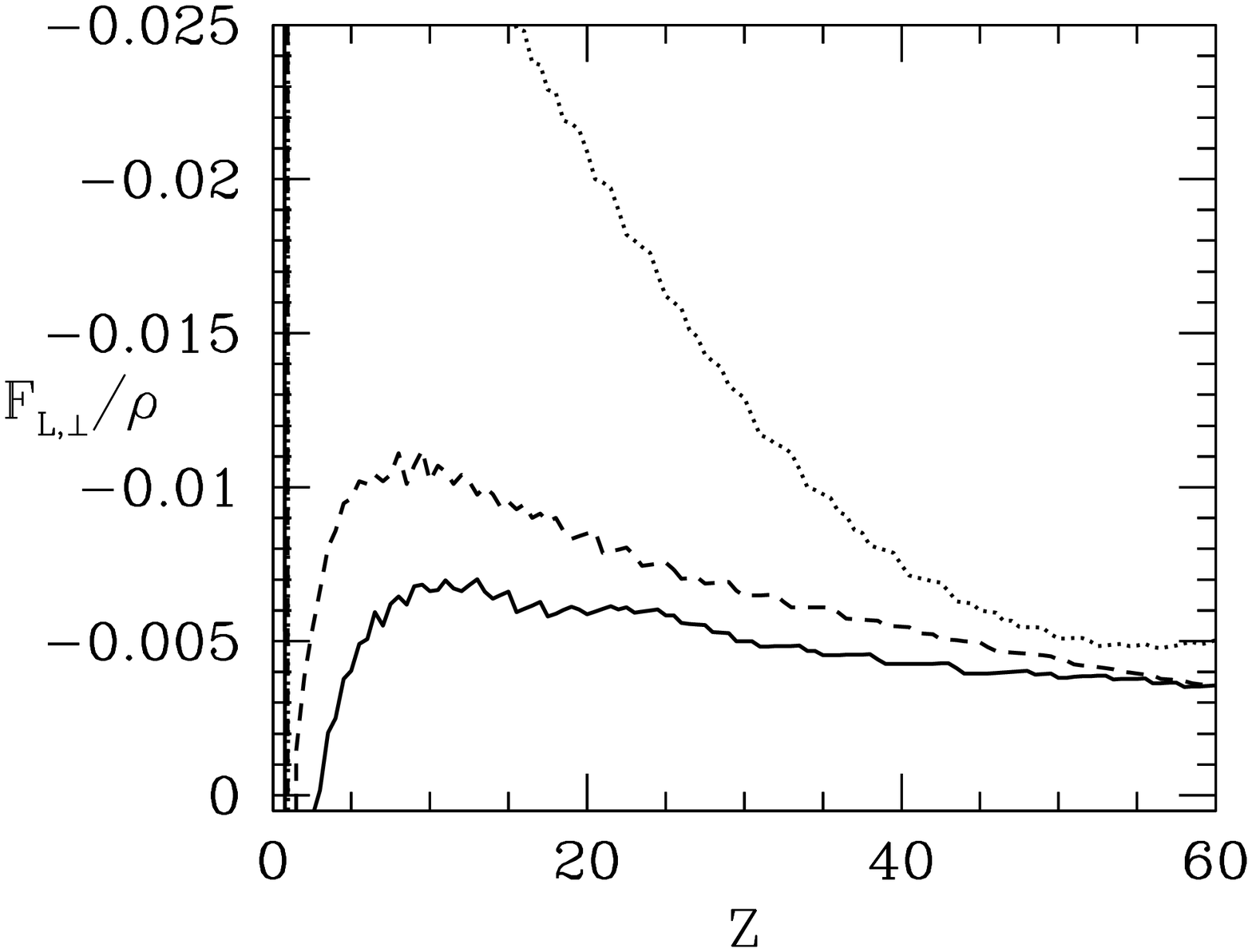}

\includegraphics[bb= 18 144 592 618,width=7cm]{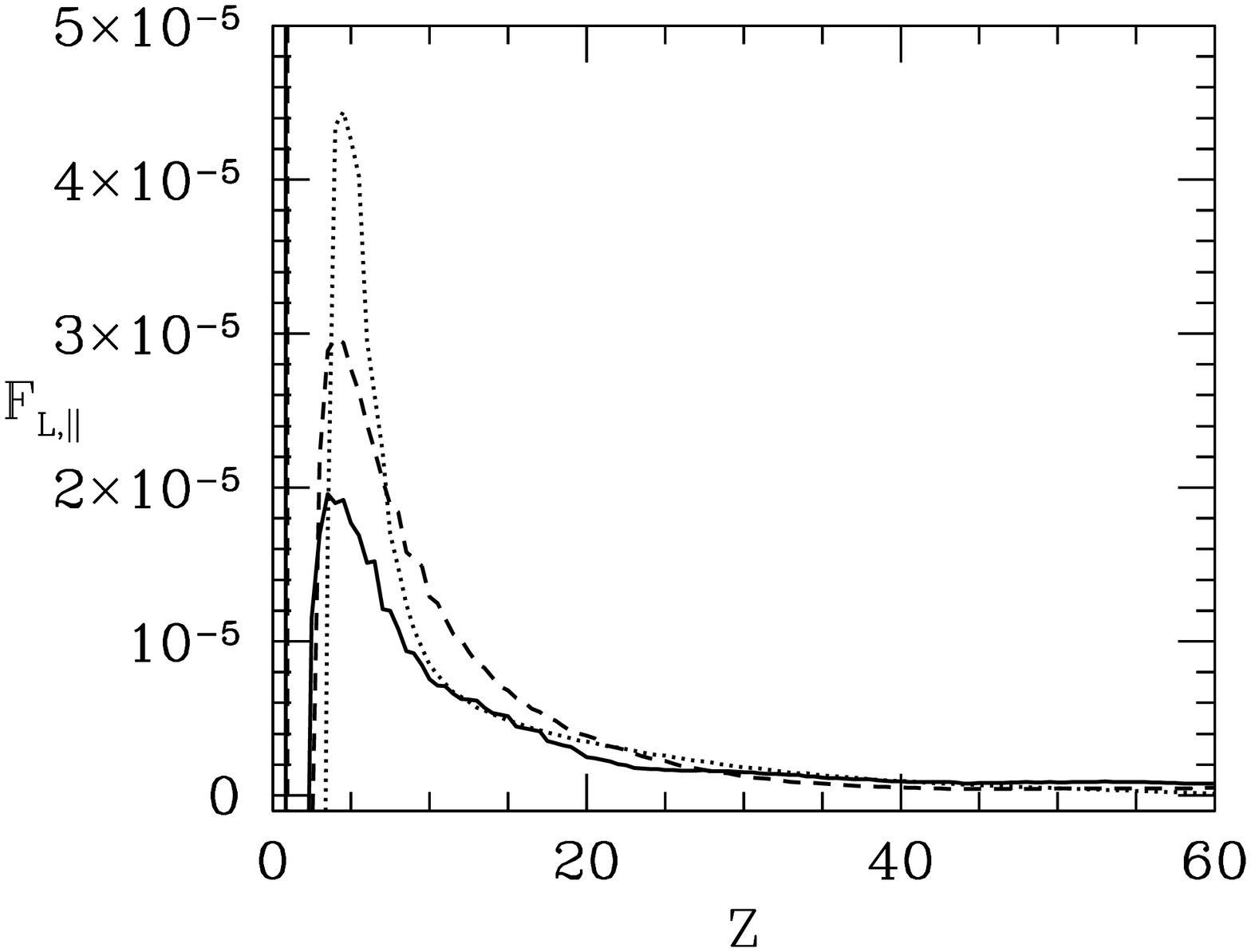}
\includegraphics[bb= 18 144 592 618,width=7cm]{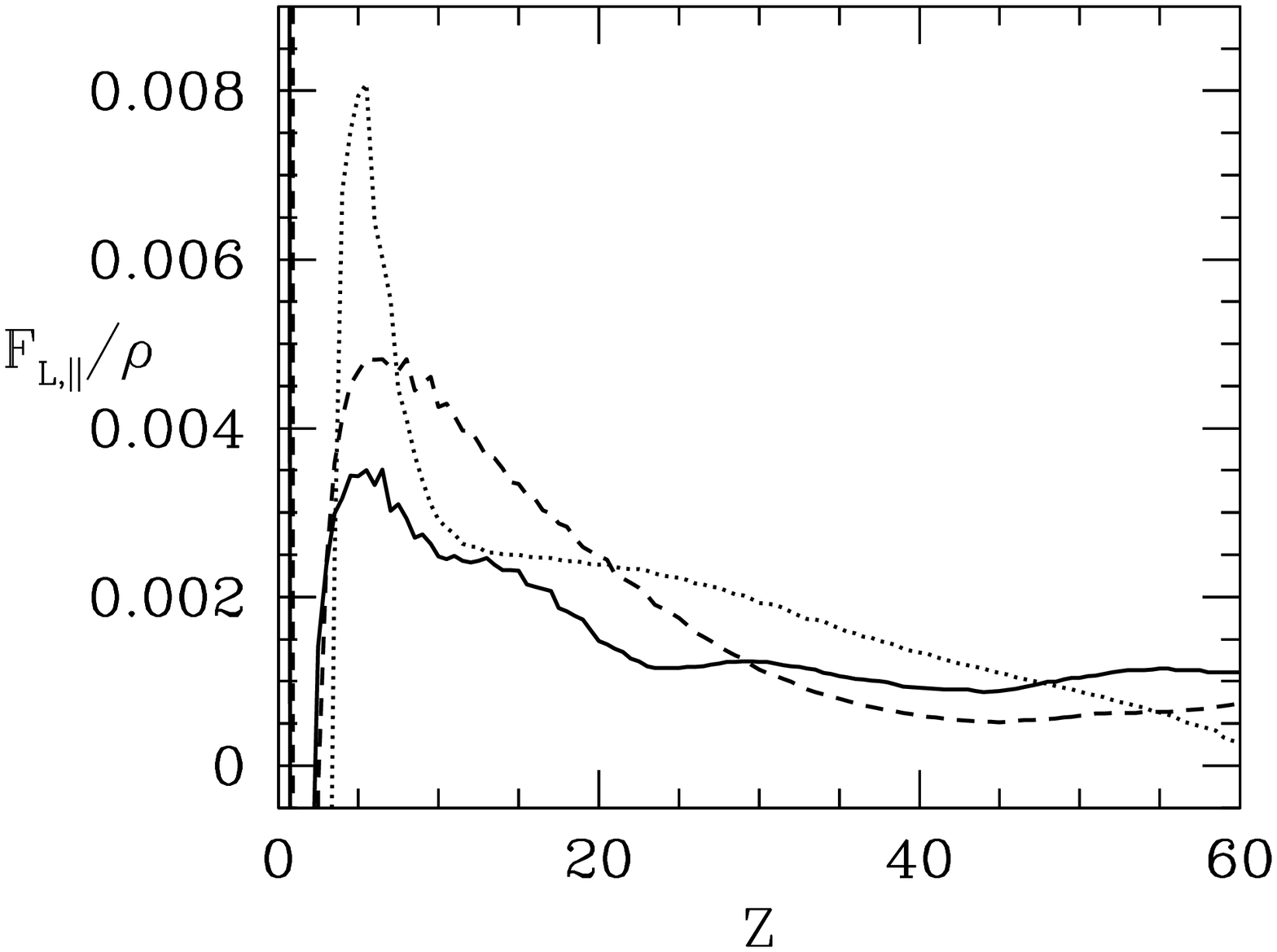}

\includegraphics[bb= 18 144 592 618,width=7cm]{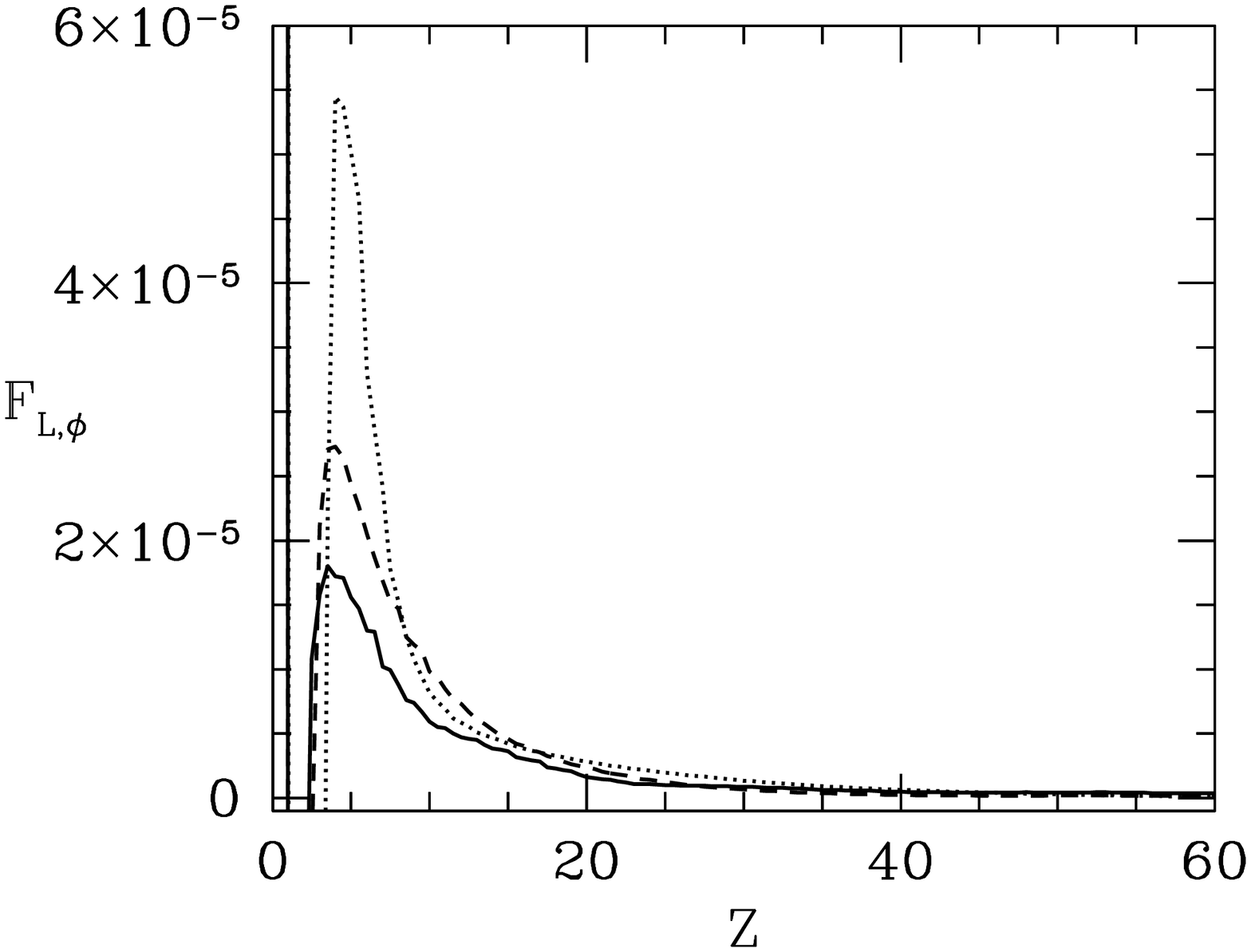}
\includegraphics[bb= 18 144 592 618,width=7cm]{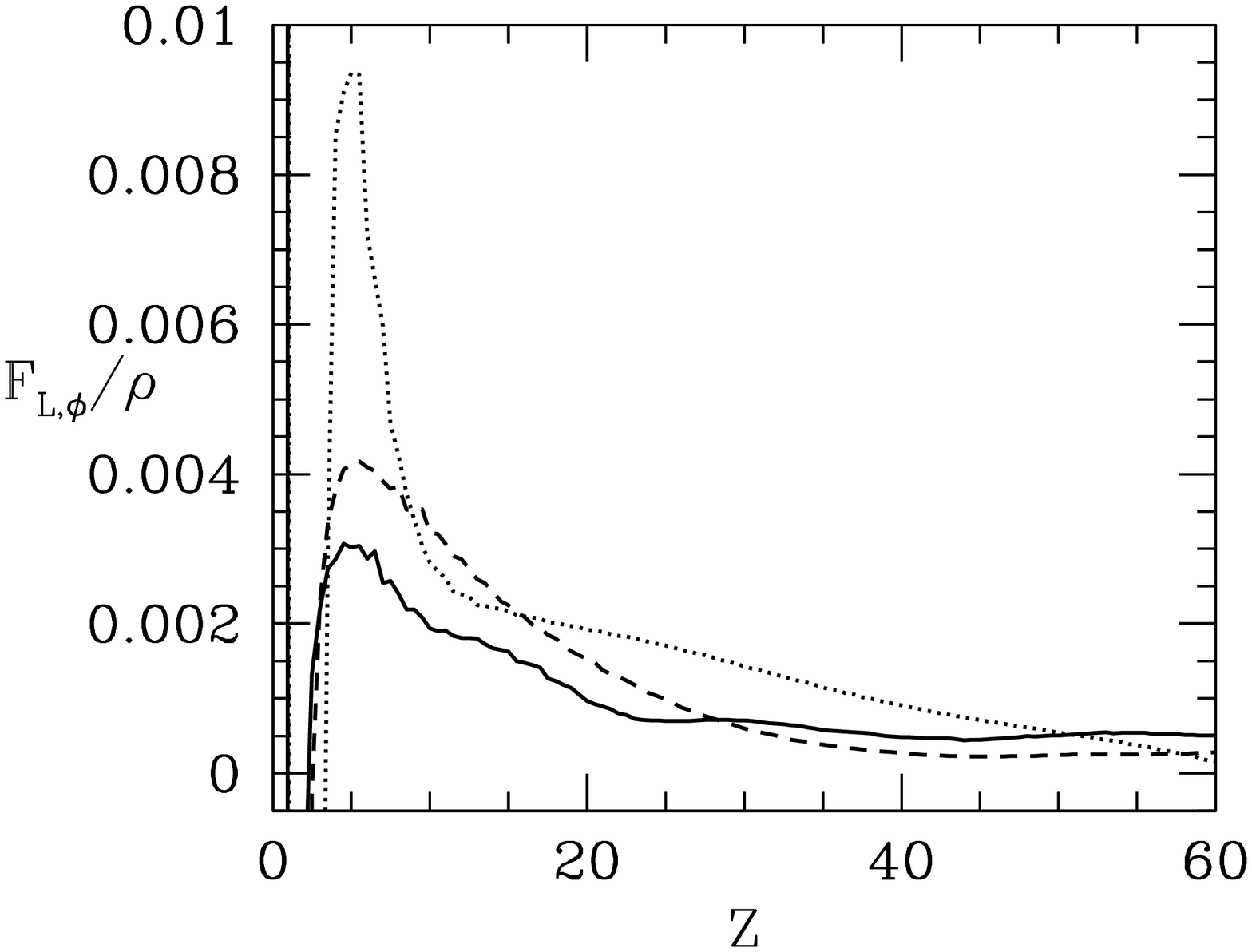}
\caption{Lorentz forces in the jet for different magnetic diffusivity
$\eta = 0, 0.1, 0.5$ ({\it solid, dashed, dotted } lines).
{\it Left}
(Normalized) values of the force component
perpendicular ({\it top}) and parallel ({\it middle}) 
to the field line and the toroidal component ({\it bottom}), 
$F_{L,\perp}$, $F_{L,\parallel}$, $F_{L,\phi}$,
along a flux surface leaving the box of the inner jet close to 
($R=20,Z=60$)-corner (see Fig.\,3 or  Fig.\,5).
For $F_{L,\perp}$ the positive sign denotes the $r$-direction
(de-collimating force).
For $F_{L,\parallel}$ the positive sign denotes the $z$-direction 
(accelerating force).
For $F_{L,\phi}$ the positive sign denotes the $\phi$-direction. 
{\it Right}
Corresponding values of the magnetic acceleration
$(F_{L,\perp}/\rho)$, $(F_{L,\parallel}/\rho)$, $(F_{L,\phi}/\rho)$.
}
\end{figure*}

\subsection{Lorentz forces in the jet}
Here, we deal with the question how the jet internal structure is 
modified by the effect of magnetic diffusivity as a result of
our numerical simulations.
Compared to the ideal MHD simulation (OP97) our results of 
  (i) a de-collimation of the poloidal magnetic field structure for any
      value of magnetic diffusivity, 
 (ii) a de-collimation of the hydrodynamic flow for strong diffusion and 
(iii) an increase of the jet velocity with increasing diffusivity 
are obtained purely by adding a physical magnetic diffusivity to the
originally ideal MHD code.

As the flow evolution in the MHD simulation results from a 
combination of various physical effects --
magnetic and inertial forces, pressure and gravity --
it is not straight forward to distinguish between these
contributions.

However, it seems to be clear that magnetic fields are the main
driver for the flow acceleration and self-collimation and that,
consequently, 
the addition of magnetic diffusion will modify the MHD structure of 
the jet.
Therefore, investigating the Lorentz forces in the quasi-stationary state 
may provide some insight into the physical mechanisms at work.

At this point it might be instructive to recall in brief the basic
mechanisms of MHD jet formation.
Following the standard model 
(e.g. Blandford \& Payne \cite{blandford}, Ferreira \cite{ferreira}),
a jet is launched as a sub-Alfv\'enic 
disk wind (by some unspecified -- but also unknown -- process)
and becomes accelerated by magneto-centrifugal forces in a strong 
poloidal magnetic field at first hand.
As the flow approaches the Alfv\'en surface, a toroidal magnetic field
component is induced (``wound-up'') due to the inertial back-reaction
of the matter on the field.
The toroidal field may lead to (de-) accelerating Lorentz forces 
$\vec{F}_{\rm L, ||} \sim \vec{j}_{\perp} \times \vec{B}_{\phi} $
and (de-) collimating forces
$\vec{F}_{L,\perp} \sim \vec{j}_{||} \times \vec{B} $ 
where, here, the perpendicular and parallel projection is made with
respect to the poloidal magnetic field 
(which, only in the case of ideal stationary MHD is parallel to the
poloidal velocity).
The toroidal Lorentz force 
$\vec{F}_{\rm L, \phi} \sim \vec{j}_{\rm p} \times \vec{B}_{\rm p} $ affects 
the angular velocity of the matter, disturbing the centrifugal balance and, 
thus, give rise also to a poloidal motion.

Therefore, a change in the jet acceleration and collimation might be explained
by the interplay of two mechanisms. 
First, the winding-up of the poloidal magnetic fields is less efficient since 
magnetic diffusion leads to a slip of matter across the field.
As a consequence, the induced toroidal magnetic field is weaker leading to
a less efficient acceleration by Lorentz forces but also to a de-collimation.
This effect applies predominantly in the super-Alfv\'enic regime.
Second, as a de-collimation of the poloidal magnetic structure also implies a 
smaller launching angle for the sub-Alfv\'enic flow, the magneto-centrifugal
acceleration mechanism may work more effective.
As a consequence, the resulting fluid velocities in the jet should be larger,
as indeed suggested by our simulations (Fig.\,7).

In Fig.\,10 we show for different magnetic diffusivity the Lorentz force
components along a field line (or, respectively, along the corresponding
magnetic flux surface) 
leaving the numerical grid of the inner jet at ($r=20,z=60$).
Note that due to the magnetic field de-collimation with $\eta$,
we compare {\em different magnetic flux surfaces}.
These flux surfaces have their foot point between $r=5$ and $r=8$ along the
disk surface
and the Alfv\'en point at about $z=25, 15, 5$ ($\eta = 0, 0.1, 0.5$).
The figure shows the Lorentz force components 
$F_{\rm L,\perp}$, $F_{\rm L,\parallel}$, $F_{\rm L,\phi}$ 
and the corresponding acceleration of the fluid $F_{\rm L}/\rho$.

The first point to mention is that the magnitude of the Lorentz force 
generally increases with increasing magnetic diffusivity.
This is interesting insofar as the magnetic field strength decreases with
increasing diffusion (Fig.\,7).
The Lorentz force has its maximum in that region {\em before} the Alfv\'en
point where the curvature of the poloidal field is largest.
Thus, {\em magnetic acceleration} mainly works in this regime.

This picture of an acceleration purely by magnetic forces is complementary
to the above mentioned picture of an enhanced magneto-centrifugal effect.
For the parallel component this may directly lead to the observed
increase in the poloidal velocity with increasing magnetic diffusivity
(see Fig.\,7).
Additionally, the higher velocity also leads to stronger inertial forces
and,
for moderate heights above the disk,
the diffusive plasma flow will tend to maintain its (radial) direction 
even if the field lines bend in direction of the jet axis.
This will re-distribute the mass flow distribution along the field line.
The parallel component decreases rapidly with increasing $z$ as it can
be expected when the jet flow becomes more and more collimated.
The same holds for the toroidal Lorentz force component. 
This component accelerates the plasma in toroidal direction leading
to an additional centrifugal effect which drives the matter in radial
direction diffusing across the magnetic field.
This is the reason for the increase of the mass flow rate
along the outer stream lines with increasing diffusivity.

While the curves for the three Lorentz force components look quite
similar at a first glimpse, we see that the corresponding components
for the {\em acceleration} are somewhat different.
The perpendicular (collimating) component\footnote{
We note that the sign for $F_{\rm L,\perp}$ is defined 
positive for the force vector pointing radially outwards.
Thus, the increase of $|F_{\rm L,\perp}|$ indicates an increase of the
collimating Lorentz force on the matter.}
of the acceleration remains on a rather high level throughout the
(inner) box. 
That means that also in the asymptotic regime of the collimated jet
these forces continue to collimate the jet flow.

On the other hand, compared to the perpendicular component,
the parallel and the $\phi$ components of the 
corresponding acceleration have a steeper maximum and decrease to
an only marginal strength beyond the Alfv\'en point.
This is what one would expect also from the standard MHD jet model.

For $\eta = 0$ and $\eta = 0.1$,
we see only a slight difference for the strength of the perpendicular
components of force and acceleration at the large distances. 
Therefore, the degree of local flow collimation should be similar,
as it is indeed visible in the poloidal {\em velocity} vectors, 
which are well aligned for the diffusivity considered (Fig.\,6).
However, we note the larger deviation of the perpendicular components for
$\eta = 0.5$, which
mirrors the fact that in this case we are above the {\em critical value}
$\eta_{\rm cr}$ concerning the mass flow collimation (Fig.\,8).

In summary, our discussion of the Lorentz forces and its associated
acceleration gives a self-consistent picture of what we have observed 
in our numerical simulation.
The perpendicular Lorentz force is essential for the collimation
throughout the entire (inner) flow.
The increase of the parallel Lorentz force for higher magnetic diffusivity
gives rise to the higher velocities in the jet flow.
The toroidal Lorentz force leads to an additional centrifugal effect
enhancing the mass flow rate in the outer (yet un-collimated) parts of the
jet flow.

%
%


\section{Summary}
In this paper we presented time--dependent simulations of the 
formation of axisymmetric protostellar MHD jets.
In particular, we were considering the effects of {\em resistive} MHD 
on the collimation and acceleration of the jet flow.
Similar to recent simulations considering the ideal MHD case 
(Ouyed \& Pudritz \cite{ouyed}, Fendt \& Elstner \cite{fendt00}), the
accretion disk has been taken as a fixed boundary condition
during the simulation, prescribing the mass flow rate and the
magnetic flux distribution.
Our initial condition is a force-free magnetic field in a hydrostatic
corona.
Our simulations were performed on a 
grid of ($z\times r$)=(280$\times$ 40) inner disk radii with 
$900 \times 200$ grid elements or on a
grid of $140 \times 40$ inner disk radii with $280 \times 80$ grid 
elements.
We find that in general the low resolution simulations were sufficient
to cover all physical effects observed in the higher resolution runs.
In our discussion we mostly concentrate on the structure of the
{\em inner jet} which is the region of $60 \times 20$ inner disk radii
close to the star.
We summarize our results as follows.

\begin{itemize}

\item[(1)]
We have successfully implemented the physical magnetic diffusivity
into the ZEUS-3D code. 
\item[(2)] 
We have discussed some analytical estimates about the strength of 
magnetic diffusivity in protostellar jets. 
We derive the distribution of magnetic diffusivity self-consistent
to the turbulent Alfv\'enic pressure which is underlying our
simulations and also present in the simulations of other authors.
Our simulations, however, are performed with a constant diffusivity.
Our main results do not depend on the actual distribution
of magnetic diffusivity.
\item[(3)] 
In the global scale of our simulation, 
the jet bow shock advances slower through the initial hydrostatic corona
for the diffusive jets.
The reason is the lower mass flux in the direction along the jet axis
in these jets.
As expected, the internal structure of the jet is less disturbed 
in the case of diffusion.
The Alfv\'{e}n surface comes closer to the disk surface.
\item[(3)] 
For our model setup we find that, similar to the case
of ideal MHD jets (Ouyed \& Pudritz \cite{ouyed}, Fendt \& Elstner \cite{fendt00}),
also resistive MHD jets can reach a quasi-stationary state.
With increasing magnetic diffusivity, the quasi-stationary
state of the jet is reached later.
\item[(4)] 
With increasing diffusivity the jet velocity increases.
The direction of the velocity vectors does, however, only
change weakly.
At the same time the poloidal magnetic field distribution 
becomes increasingly de-collimated.
\item[(5)] 
As a proper measure of the degree of collimation we suggest the
mass flux.
If we compare the mass flow rates through the grid boundaries for
different diffusivity we find strong indication for the existence
of a {\em critical value} for the magnetic diffusivity $\eta_{\rm cr}$
concerning the jet collimation.
Beyond this value we still find an almost cylindrically  collimated 
stream along the jet axis, however, the bulk mass flow is in radial
direction. 
For our setup, the critical (normalized) diffusivity is about
$\eta_{\rm cr}\approx 0.3$.
\item[(6)] 
We discuss a self-consistent picture where these effects
of jet de-collimation and acceleration are explained in the
context of Lorentz forces.
The perpendicular Lorentz force is essential for the collimation
throughout the entire flow along the jet axis.
The parallel Lorentz force increases for increasing magnetic diffusivity
and gives rise to the higher velocities in the jet flow.
The toroidal Lorentz force accelerates the plasma in toroidal direction.
This leads to additional centrifugal forces re-distributing the mass flow
rates across the magnetic flux surfaces towards the outer (yet
un-collimated) parts of the flow.
The latter two components play no role for larger distances along the 
flow.

\end{itemize}

With our results we have shown that magnetic diffusivity plays indeed a role
for the jet formation process.
Turbulence as a natural (and necessary) property of accretion disks will
naturally enter the disk wind and will be further advected into the jet.
As we see in our simulations only a weak collimation for a high magnetic 
diffusivity, a hypothetical, and for sure exaggerated, conclusion might
therefore be that highly turbulent disks cannot drive a collimated jet
mass flow.
Such a claim may eventually be tested by the astronomical observation
and may also give some hint to answer the question why some disks have jets
and some do not.

Our present study should be understood as a first step in the right 
direction. 
Future work may improve the numerical resolution and the grid size
but may also consider e.g. an additional central stellar magnetosphere
as a boundary condition.
The most interesting (but also most difficult) prospect would be to
include the evolution of the disk structure in the simulation.

\begin{appendix}

\section{Numerical Tests}
As we have introduced the effect of physical magnetic diffusivity into the
ideal MHD ZEUS--3D code,
careful tests were necessary to prove our implementation.
In particular we checked the time scales introduced by magnetic diffusion
and the behavior along the boundaries.
The boundary conditions for an axisymmetric jet (``outflow'', ``inflow ''and
``reflecting'') are quite different from what is e.g. needed in box simulations
used for other scientific questions.

We defined two numerical tests for our diffusive code.
In both cases the code basically solves the diffusion equation.
We obtained this limit by setting the initial density in the simulation
to arbitrarily large values (here the normalized $\rho \simeq 10^9$) 
effectively reducing any fluid motions in our simulations.
The first test example is the analytical solution of the diffusion
equation in Cartesian coordinates, 
the second example an axisymmetric torus of purely toroidal magnetic field 
in cylindrical coordinates.

\subsection{Analytical solution to the diffusion equation}
In Cartesian coordinates $ (x,y,z)$ the solution of the 
one dimensional diffusion equation for infinite space is
\begin{equation}
B_z(y,t)=\frac{1}{\sqrt{t}}\exp\left({-\frac{(y-y_0)^2}{4\eta t}}\right)\ ,
\end{equation}
with the magnetic diffusivity $\eta=c^2/(4\pi\sigma )$. 

As a test for our code, we choose as initial condition the magnetic
field $B_z(x,y) = B_z(y)$ for a certain time $t=t_0$ from eq.\,(A.1).
For the two-dimensional numerical grid we prescribe ``free'' (i.e. outflow)
boundary conditions in $x$-direction and a time-varying field for the
boundaries in $y$-direction.

Figure A.1 show the result of our simulations for the time steps
$t=t_0 \Delta t$ ($\Delta t = 0.1,0.2,0.3,0.4$) 
for a magnetic diffusivity $\eta=1.0$ in
comparison with the analytical results.
As result, we obtain a perfect agreement between the numerical simulation
and the analytical solution.

\setlength{\unitlength}{1mm}
\begin{figure}
\includegraphics[bb= 18 144 592 718,width=4.3cm]{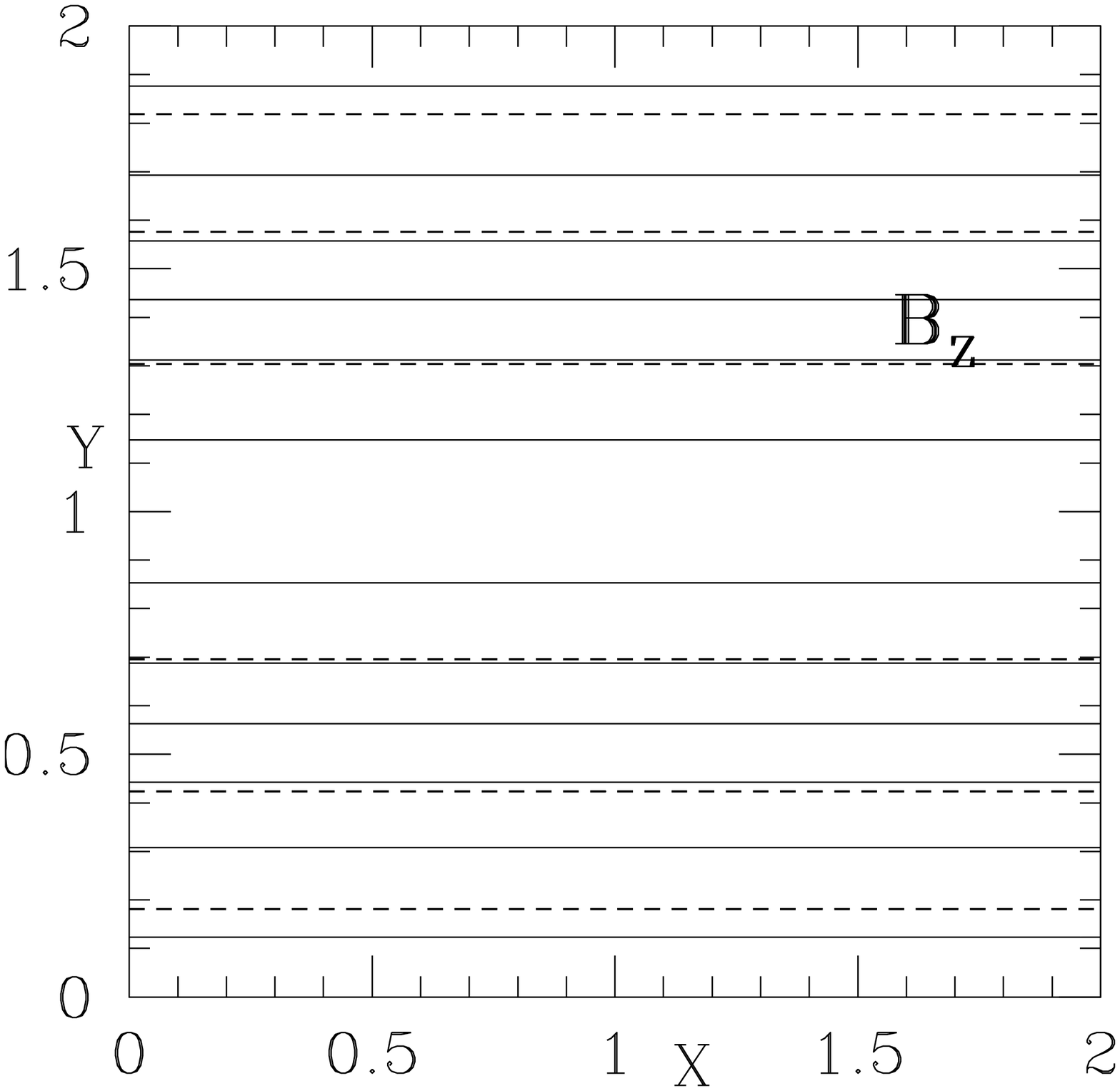}
\includegraphics[bb= 18 144 592 718,width=4.3cm]{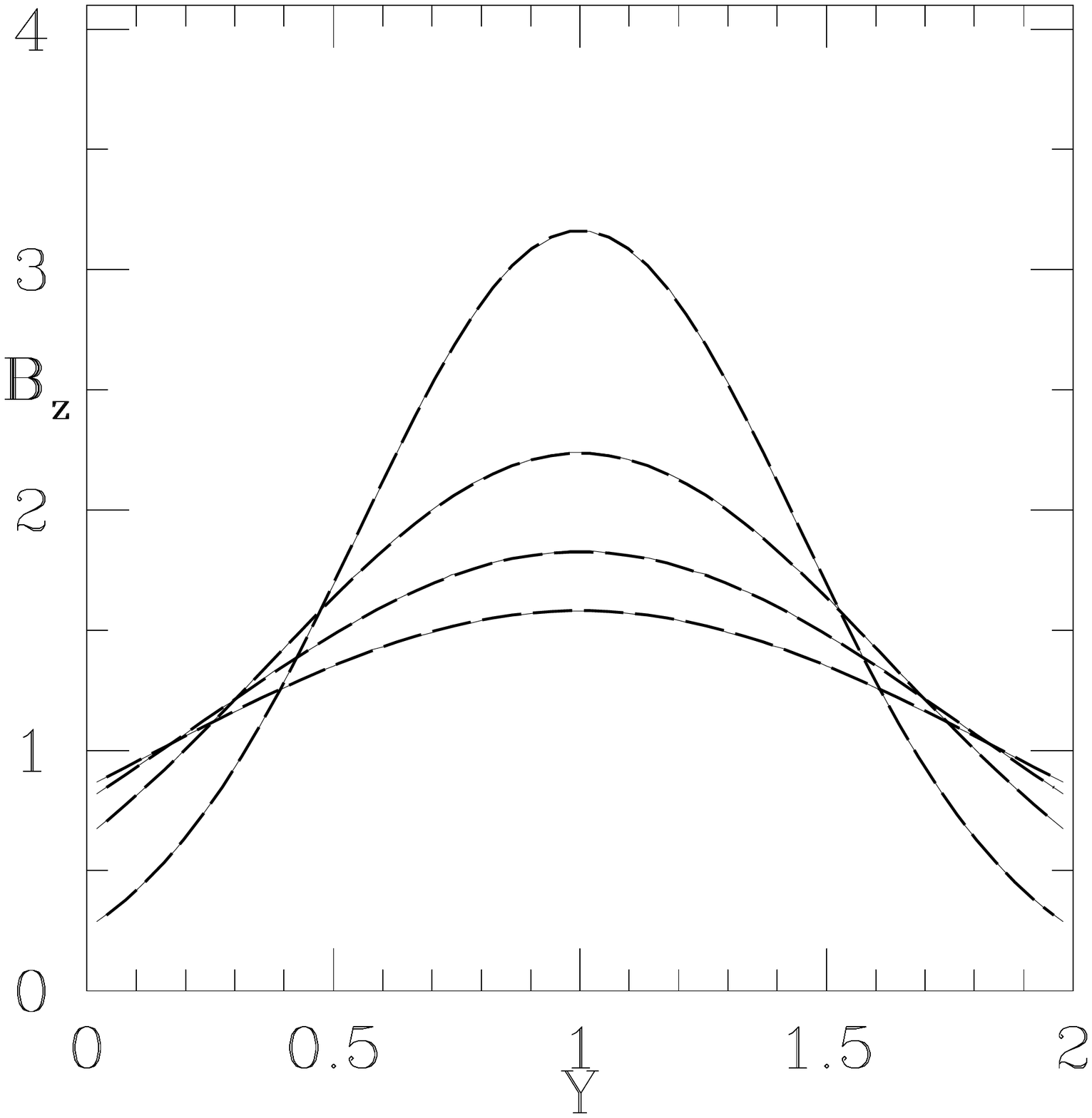}
\caption{Numerical test of magnetic diffusion. 
Grid size $50 \times 50$
elements for a normalized physical grid $2.0\times 2.0$.
{\it Left} 
Isocontours of the magnetic field strength $ B_{\rm z}(x,y)$ (normalized
units) for different time steps $ t= t_0 + \Delta t$,
$\Delta t = 0.0$ ({\it solid line}), 
$\Delta t = 0.1$ ({\it dashed line}).
{\it Right} 
Normalized intensity profile of the magnetic field strength across the
two-dimensional box along $x=1$ for different time steps $ t= t_0 + \Delta t$
with $\Delta t = 0.0, 0.1, 0.2, 0.3$ ({\it top} to {\it bottom} curve).
Comparison between the analytical solution {\it solid lines} and the 
numerical simulation ({\it dashed lines}).
}
\end{figure}

\subsection{Toroidal field torus}
Here, our aim is to check how our code treats magnetic diffusion in
cylindrical coordinates, along the outflow boundary in $r$-direction
and along the symmetry axis.
As initial condition, we define a torus of toroidal magnetic field 
\begin{equation}
B_\varphi(r,z,t_0=0.1)=\frac{1}{t_0} 
\exp\left({-\frac{(r-r_0)^2+(z-z_0)^2}{4\eta t_0}}\right)\ .
\end{equation}
Figure A.2 show the result of our simulations for the time steps
$t=t_0 \Delta t$ ($\Delta t = 0.1,0.2,0.3,0.4$)
for a magnetic diffusivity $\eta=1.0$ in
comparison with the analytical results.
The simulation shows how the peak of the field distribution 
moves slightly inwards from its initial central position as the
field diffuses.
Note that as the field diffuses outwards the volume over which the 
toroidal field is distributed increases.
Therefore the decrease in field strength in outward direction.
Along the symmetry axis the field strength remains zero,
whereas the field strength along the outflow boundaries increases.
No boundary condition is prescribed here.

Although there is no analytical solution to compare with, this
simulation gives again convincing evidence that we properly
incorporated the  magnetic diffusion in the ZEUS-3D code.

\setlength{\unitlength}{1mm}
\begin{figure}
\parbox{100mm}
\thicklines
\epsfysize=44mm
\includegraphics[bb= 18 144 592 718,width=4.3cm]{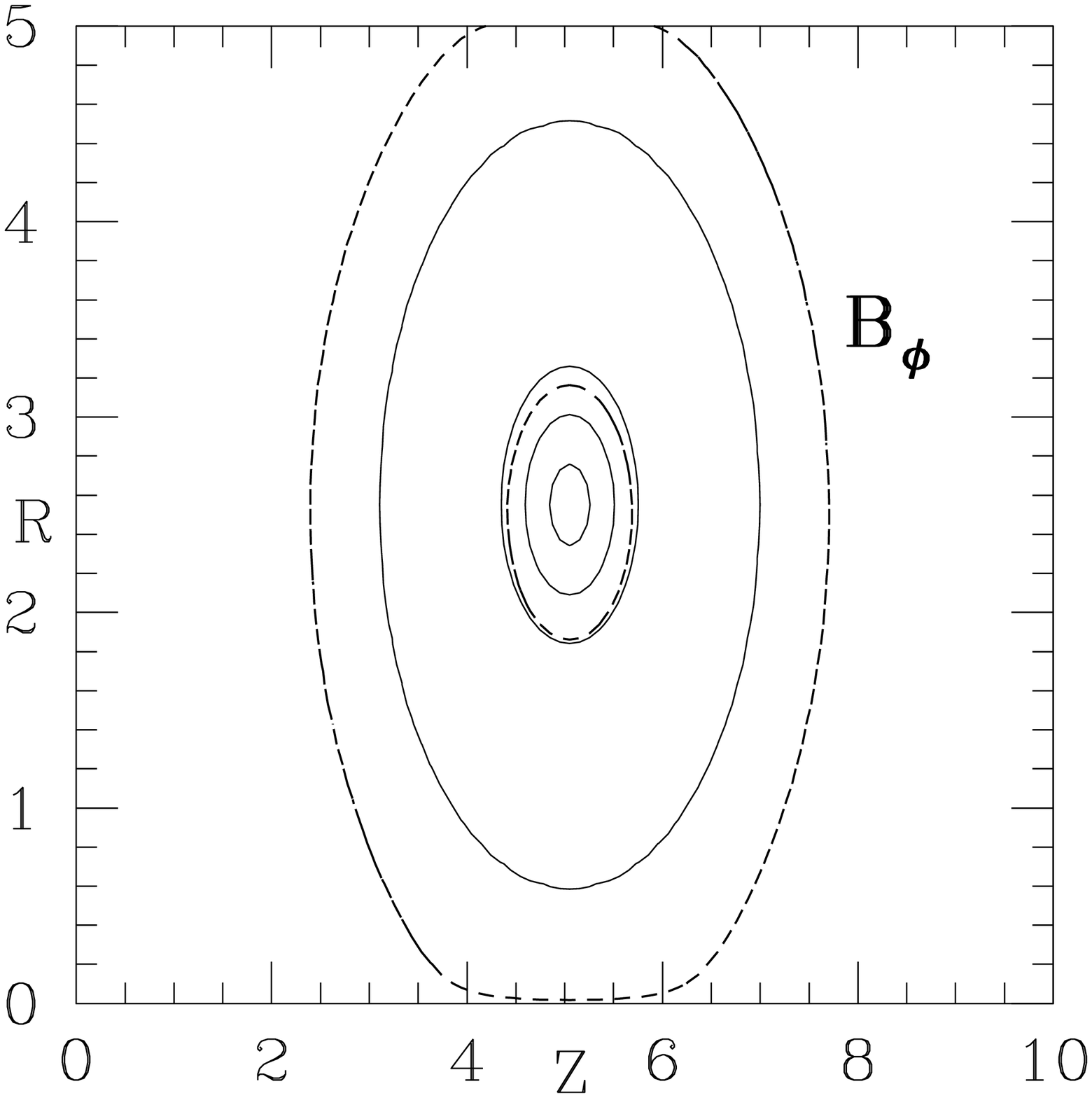}
\includegraphics[bb= 18 144 592 718,width=4.3cm]{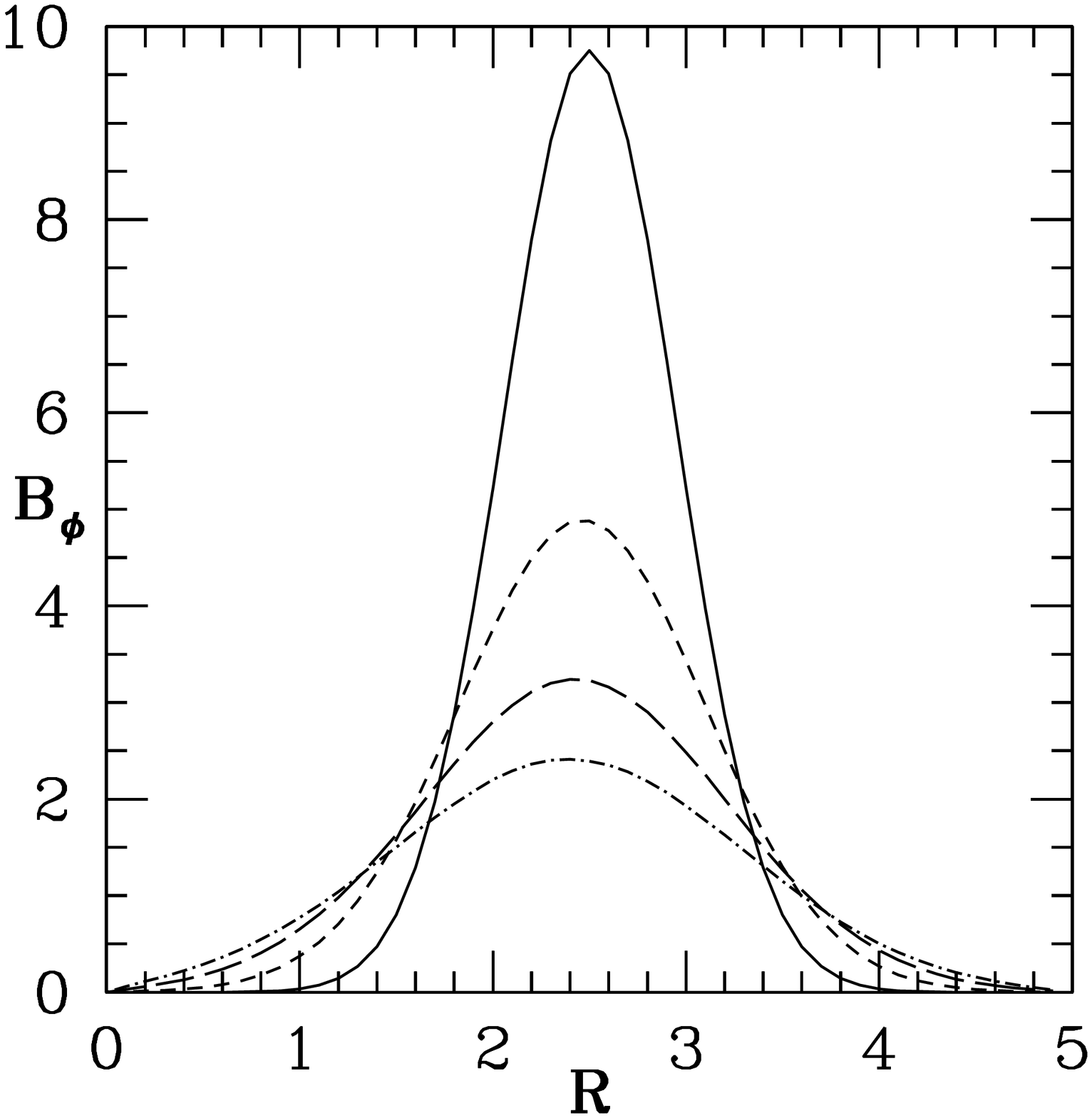}
\caption{Numerical test of magnetic diffusion in cylindrical coordinates. 
Grid size $100 \times 50$
elements for a normalized physical grid $10.0\times 5.0$.
{\it Left} 
Isocontours of the toroidal magnetic field strength for 
different time steps $ t= t_0 + \Delta t$,
$\Delta t = 0.0$ ({\it solid line}),
$\Delta t = 0.1$ ({\it dashed line}).
{\it Right} 
Normalized intensity profile of the magnetic field strength across the
two-dimensional box along $z=5$ for different time steps $ t= t_0 + \Delta t$
with $\Delta t = 0.0, 0.1, 0.2, 0.3$ ({\it top} to {\it bottom} curve).
}
\end{figure}

\section{Time scales in the simulation}
Table 1 shows the time scales and typical number values applied in our
jet simulations for different magnetic diffusivity $\eta$.
As an example we refer to the simulation runs with relatively low
numerical resolution.
The diffusive time step (i.e. the local time scale for magnetic diffusion)
is $\tau_{\eta} = (\Delta x/\eta^2)$ with the size of the grid element
$\Delta x$.
For comparison we show the Alfv\'en time step 
$\tau_{A} = (\Delta x/v_{\rm A})$.
The other two numbers consider the global evolution of the jet flow.
For the magnetic Reynolds number $R_{\rm m} = (v_{\rm A} r_{\rm max}/\eta)$
we have considered the global size of the jet and similar
for the magnetic turbulence parameter $\alpha_{\rm m}$.

\begin{table}
\caption{Typical numbers for our simulations for different magnetic 
diffusivity $\eta$.
Diffusive time step $\tau_{\eta}$, Alfv\'en time step $\tau_{A}$, 
the global magnetic Reynolds number $R_{\rm m}$ 
and the magnetic turbulence parameter $\alpha_{\rm m}$.}
\hspace{1.5cm}\begin{tabular}{|c||c|c||c|c|c|}
\hline
$\eta$ & $\tau_{\eta}$ & $\tau_{A}$ & $R_m$ & $\alpha_m$\\
\hline
\ \  0 & $ \infty $ & 0.258 & $\infty$ & 0 
\\
0.1 & 0.625 & 0.293 & 40 & 0.025 
\\
0.15 & 0.417 & 0.084 & 30 & 0.033 
\\
0.25 & 0.25 & 0.136 & 20 & 0.05 
\\
0.5 & 0.125 & 0.080 & 10 & 0.1
\\
1 & 0.063 & 0.183 & 8  & 0.125 
\\
2.5 & 0.025 & 1.394 & 4  & 0.25 \\
\hline
\end{tabular}
\end{table}

\end{appendix}
 
\begin{acknowledgements}
We thank the LCA team and M.~Norman for the possibility to use the
ZEUS-3D code. 
We acknowledge helpful discussions with D.~Elstner, U.~Ziegler
and R.~Ouyed.
This work was partly financed by the German science foundation
in the DFG Schwerpunktprogramm ``Physik der Sternentstehung''
(FE490/2-1). 
The referee, Takahiro Kudoh, is acknowledged for useful advice 
concerning the presentation of our results and for drawing our 
attention to some very recent papers in the literature.
\end{acknowledgements}


\end{document}